\documentclass[11pt]{article}

\usepackage{amsmath,amssymb}
\usepackage[paper=a4paper,text={17cm,25cm},centering]{geometry}
\usepackage{graphicx}
\usepackage{ulem}
\usepackage{rotating}
\usepackage{enumerate,mdwlist}

\def\be{\begin{equation}}
\def\ee{\end{equation}}
\def\Xmin{{X_{\text{min}}}}

\begin{document}

\title{Monte Carlo study of the tip region of branching random walks evolved
  to large times
}

\author{
  Anh Dung Le${}^{(1)}$,
  Alfred H. Mueller${}^{(2)}$,
  St\'ephane Munier${}^{(1)}$\vspace{1em}\\
    \footnotesize\it (1) CPHT, CNRS, \'Ecole polytechnique,
  IP Paris, F-91128 Palaiseau, France\\
    \footnotesize\it (2) Department of Physics, Columbia University,
  New York, NY 10027, USA\\
}
\date{
  \vskip 1em
%March 9, 2020}
\today}

\maketitle

\begin{abstract}
  We implement a discretization of the one-dimensional
  branching Brownian motion in the form of a Monte Carlo event generator,
  designed to efficiently produce ensembles of realizations in which the rightmost
  lead particle at the final time~$T$ is constrained to have a position $X$
  larger than some predefined value $X_{\text{min}}$. The latter may be chosen
  arbitrarily far from the expectation value of $X$, and the evolution
  time after which observables on the particle density near the
  lead particle are measured may be as large as $T\sim 10^4$.
  We then calculate numerically
  the probability distribution $p_n(\Delta x)$ of the number $n$ of particles
  in the interval $[X-\Delta x,X]$ as a function of $\Delta x$.
  When $\Xmin$ is significantly smaller than the expectation value of the
  position of the rightmost lead particle,
  i.e. when $X$ is effectively unconstrained, we check that
  both the mean and the typical values of $n$
  grow exponentially with $\Delta x$, up to a linear prefactor
  and to finite-$T$ corrections.
  When $X_{\text{min}}$ is picked far ahead of the latter
  but within a region extending over a size of
  order $\sqrt{T}$ to its right, the mean value of the
  particle number still grows exponentially with $\Delta x$,
  but its typical value is lower by a multiplicative factor consistent with
  $e^{-\zeta\Delta x^{2/3}}$, where $\zeta$ is a number of order unity.
  These numerical results bring strong support
  to recent analytical calculations and conjectures in the
  infinite-time limit.
\end{abstract}

\section{Introduction}

Understanding correlations of particles generated by
one-dimensional branching random walks is an outstanding general
problem in mathematics, with numerous potential applications in
different fields of science.
%physics, biology, computer science.
Various observables sensitive to these correlations may be derived
from solutions to the Fisher-Kolmogorov-Petrovsky-Piscounov (FKPP)
equation \cite{f,kpp} when the underlying
stochastic process is the branching Brownian motion,
or to an equation in the same universality class
when the underlying process is some
branching random walk.

The FKPP equation is relevant to many
contexts (for a review, see e.g.~\cite{VANSAARLOOS200329}).
In biology, it may describe the
spread of a gene or of a disease in a population: This is
the field in which it was first written down in the literature~\cite{f,kpp}.
In chemistry, it stems from a mean-field approximation
for reaction-diffusion processes.
In computer science, it appears in the study of the statistics of the
heights of search trees~\cite{Majumdar_2005}. In physics, 
it was also shown to be relevant for quite unexpected problems:
For example, to understand 
the mean-field theory of disordered systems such as
spin glasses or directed polymers
in random media~\cite{derrida1988polymers}, and
to describe some properties of scattering processes
in elementary particle physics (for a review, see e.g.~\cite{Munier:2009pc}).
The latter was the initial motivation of the particle physicists
for studying the FKPP equation~\cite{Munier:2003vc}.

Of particular interest here is the fact that the FKPP equation
(or, more generally, FKPP-like equations)
encodes the time evolution of generating functions
of particle number
probabilities $p_n(\Delta x)$ in intervals spanning a fixed
distance $\Delta x$ to the left of
the position $X$ of the rightmost particle in realizations of
the branching Brownian motion (or, more generally, of branching random walks)
\cite{Brunet2011}.
So the problem of understanding these distributions of
particles may be formulated elegantly in terms of
equations in the universality class of the FKPP equation.
Therefore, it essentially boils down to
the well-defined mathematical problem of
solving nonlinear partial differential equations.
But since FKPP-like equations do not have complete analytical solutions,
finding expressions for the observables of interest is not an easy task.
On the other hand, these equations can be integrated numerically
quite easily. However, extracting from these data either
the probabilities $p_n(\Delta x)$ or their moments
is in general not doable in practice, except for small values of $n$
and low-lying moments.

Recently, an analytical study of the generating function
of $p_n(\Delta x)$ in the asymptotic limit of large $T$,
large interval size $\Delta x$, and large distance $X-m_T$
between the position $X$ of the tip and its expectation value
$m_T$ led to a few new results:
The generating function was shown to exhibit a peculiar scaling form,
and a conjecture for the typical value of $n$ as a function of $\Delta x$
was formulated~\cite{Mueller:2019ror}.
While the scaling of the generating function could
be checked by numerical integration of the relevant FKPP equation,
arriving at a numerical determination of $p_n(\Delta x)$ 
requires a Monte Carlo implementation of the stochastic process.
However, this is not straightforward, since the configurations
of interest are very rare
realizations of branching random walks evolved to large times.
A naive implementation is doomed to fail. Algorithms to generate ensembles
of rare events have been proposed in other contexts,
see e.g.~\cite{Giardina2011}.
They would not directly apply to the problem we are addressing here, but
as will be shown below, it is possible to design one specifically.

The main thrust of the present work is to introduce an algorithm
for generating these rare realizations, and to apply an implementation
of it to the numerical study of $p_n(\Delta x)$
as a function of the size of the
interval $\Delta x$. The numerical output is eventually
confronted to the analytical expectations.

The principle of the algorithm is explained in Sec.~\ref{sec:principle}
in the case of the branching Brownian motion, and then of a branching random
walk discrete in space and time, which is the process we implement
numerically. The numerical results
are presented in the subsequent Sec.~\ref{sec:results}, after a quick study
of the model we have actually implemented and after a short review of the
known analytical results and of the conjectures to be tested.

%%%%%%%%%%%%%%%%

\section{\label{sec:principle}Generating realizations
conditioned to a minimum position of the rightmost lead particle}

For the sake of exposing the principle of the
algorithm as simply as possible, we are first going to address
the branching Brownian motion (BBM) process in one dimension.
We shall then introduce a model of a branching random walk
discrete in space and time, which is
a discretization of the latter, that proves
more convenient to implement in the form of a computer code.

\subsection{Branching Brownian motion}

The BBM of a set of particles is defined by
two elementary processes: Each particle evolves in time independently
of all the other particles through
a Brownian motion in space, until it reaches the final time~$T$, or until it
is randomly replaced by two particles at its current position. For definiteness,
we shall fix the diffusion constant to $\frac12$, and the branching rate to unity.
After a branching has occured, the offspring evolve further
independently through the
two same elementary processes (namely diffusion and branching).

A Monte Carlo implementation of the BBM is in principle straightforward,
but since the number of particles in the system grows on the average
exponentially with time, it is difficult to generate ensembles
of realizations for times larger than typically~$T\sim 10$:
Indeed, with a continuous process, each particle
must be followed individually.
Furthermore, for our purpose which is to focus on rare
events in which the lead particle is very far from its expected
position, a naive implementation of the elementary
processes is unpractical.

Specifically, we want to generate ensembles of realizations of the BBM
which, at a (large) time $T$,
contain at least one particle located at a position~$X$ not smaller than
some fixed $\Xmin$. The main idea of the algorithm
is to mark the first particle
from the beginning of the evolution, distinguishing it by conditioning
its trajectory to arriving 
at a position in the interval $[\Xmin,+\infty[$
    at time $T$. This is illustrated in Fig.~\ref{fig:scheme}.
A similar idea was proposed in Ref.~\cite{aikedon2013}, but it
was not used for the purpose of numerical calculations.

\subsubsection{Probability distribution
  of the first branching of the marked particle}

We start with a single particle at time $t$, which we put at position $x$.
We will need the probability $Q$ that there is no
particle at or to the right of the position
$x+\xi$ at time $t+\tau$ ($\xi$ may be any real number, $\tau$
is nonnegative) in the set
generated by the BBM.
It is well-known that it solves the FKPP equation\footnote{%
  We will call ``FKPP equation'' either Eq.~(\ref{eq:FKPP_Q}),
  or the equivalent equation for the function $1-Q$.
  Note that various equations in the same universality class are called after
  Fisher and Kolmogorov et al, with different values of the diffusion constant and
  branching rate, and, sometimes,
  different forms for the nonlinear term.
  }
\be
\partial_\tau Q(\tau,\xi)=\frac12\partial_\xi^2 Q(\tau,\xi)
-Q(\tau,\xi)+ Q^2(\tau,\xi)
\label{eq:FKPP_Q}
\ee
with, as an initial condition, 
$Q(\tau=0,\xi)=1-\Theta(-\xi)$.

We now write down the probability density that the first splitting occurs
at time $t_1\geq t$ at position $x_1$ and that there is at least
one particle with position $x_T\geq \Xmin$ in the ensemble at time~$T$.
This probability density writes as the sum of two terms:
\begin{multline}
\pi_{T-t,\Xmin-x}(t_1-t,x_1-x)=\Theta(T-t_1)\times e^{t-t_1}g(t_1-t,x_1-x)\left[
  1-Q^2(T-t_1,\Xmin-x_1)\right]\\
+\left[1-\Theta(T-t_1)\right]\times
e^{t-t_1}\int_{\Xmin}^\infty
d\tilde x_T\,g(T-t,\tilde x_T-x)\times g(t_1-T,x_1-\tilde x_T),
\label{eq:pi}
\end{multline}
where the function
\be
g(\tau,\xi)\equiv\frac{1}{\sqrt{2\pi \tau}} e^{-\xi^2/2\tau}
\ee
is the distribution of the position $\xi$ at time $\tau$
of a particle that undergoes a pure Brownian motion with diffusion constant $\frac12$
starting from $\xi=0$ at the initial time $\tau=0$:
It simply solves the diffusion equation
$\partial_\tau g(\tau,\xi)=\frac12\partial_\xi^2 g(\tau,\xi)$ with the initial
condition $g(\tau=0,\xi)=\delta(\xi)$.

The first term in Eq.~(\ref{eq:pi}), that we shall denote by
$\pi^{(1)}_{T-t,\Xmin-x}(t_1-t,x_1-x)$
represents the contribution of the case
in which the first splitting occurs before or at time $T$.
That is to say, it is the joint probability density of
the space-time coordinates $(t_1,x_1)$ of the first branching point,
conditioned to the fact that the marked particle 
arrives at a position not smaller than $\Xmin$ at time~$T$, and that
a branching of this marked particle occurs before or at time $T$.

The second term takes into account the case in which there
is no splitting before $T$.
In that case,
we are actually not interested in the values of
$t_1$ and $x_1$. We shall denote by $\pi^{(0)}_{T-t,\Xmin-x}$ this term
marginalized with respect to the variables~$(t_1,x_1)$.
It simplifies to
\be
\pi^{(0)}_{T-t,\Xmin-x}=e^{t-T} \int_\Xmin^\infty d\tilde x_T\, g(T-t,\tilde x_T-x)
=\frac12 e^{t-T}\text{erfc}\left(\frac{\Xmin-x}{\sqrt{2(T-t)}}\right).
\ee
This is obviously interpreted as the joint probability
that the particle does not split before $T$,
and that it passes through a position not smaller than $\Xmin$ at time $T$.

We check that the integral of $\pi_{T-t,\Xmin-x}(t_1-t,x_1-x)$
over $t_1$ and $x_1$ is nothing but the probability that
there is at least one particle in the set with position contained
in the interval $[\Xmin,+\infty[$
at time~$T$, that is to say, the solution to the FKPP equation~(\ref{eq:FKPP_Q}):
\be
\int_t^T dt_1\int_{-\infty}^{+\infty}dx_1\,\pi^{(1)}_{T-t,\Xmin-x}(t_1-t,x_1-x)+
\pi^{(0)}_{T-t,\Xmin-x}=1-Q(T-t,\Xmin-x).
\ee

\subsubsection{Description of the algorithm}

The algorithm for generating realizations of sets of particles which contain
at least one particle with position not smaller than~$\Xmin$ at time~$T$
starts with the determination of the space-time coordinates of
the successive branchings of the marked particle.
One first decides whether the next branching occurs
before the time $T$, knowing that the probability that this happens reads
\be
1-\frac{\pi^{(0)}_{T-t,\Xmin-x}}{1-Q(T-t,\Xmin-x)}.
\ee
If the marked particle does not branch before $T$,
its final position $x_T$, at time $T$,
can be generated according to the probability density
\be
\frac{e^{t-T}g(T-t,x_T-x)\Theta(x_T-\Xmin)}{\pi^{(0)}_{T-t,\Xmin-x}}.
\ee
If, instead, it branches in the time interval $[t,T]$,
the space-time coordinates of the
splitting point $(t_1,x_1)$
are drawn according to the probability density
\be
\frac{\pi^{(1)}_{T-t,\Xmin-x}(t_1-t,x_1-x)}{
  1-Q(T-t,\Xmin-x)-\pi^{(0)}_{T-t,\Xmin-x}}.
\ee
Note that this assumes the knowledge of the probability $\pi^{(0)}$
and of the density $\pi^{(1)}$,
in principle in the whole time range $[0,T]$ and for all values
of the spatial variable.
There is no analytical expression for these quantities:
They must be computed by integrating numerically the FKPP equation.
This requirement
actually sets the main limitation on the numerical calculation:
One needs to be able
to produce the value of this probability distribution
at each evolution step of the marked particle at least
(in the simplest algorithm), with a
very good accuracy. This limitation is particularly challenging to overcome
in the case of a continuous process in space and time such as the BBM.

Once the branching point $(t_1,x_1)$ has been chosen, the marked particle is any
of the two identical particles which stem from the splitting.
Its next branchings $(t_2,x_2),\cdots,(t_l,x_l)$ and eventually its position $x_T$
at the final time $T$ are generated by iterating the procedure just described:
It is enough to go through the same steps as above, after replacement of
$(t,x)$ by $(t_1,x_1)$, and so on.

Once the marked particle has been evolved all the way to time $T$,
its offspring is
evolved by independent branching Brownian motions, with or without any
conditioning, starting from the set of points
$\left\{(t_i,x_i);i\in\{1,\cdots,l\}\right\}$ until the time $T$.
In the case in which one wants to get complete realizations at
time $T$, these BBMs are not conditioned.
Instead, in the case 
in which one is only interested in the particle content
at the final time $T$
in an interval near the tip, that is to say, in particles which have
positions not less than
$\Xmin-\Delta x$, then the generic offspring $i$ born at time $t_i$
at position $x_i$ is evolved only if it
has at least one such particle in its descendence at time~$T$:
This happens with probability $1-Q(T-t_i,\Xmin-\Delta x - x_i)$.
If it is evolved, then we mark it, and its descendence is generated 
by recursive iteration of the algorithm.

Finally, if one wishes to output the particle content of the realization
at intermediate times, it is enough to generate realizations of the
Brownian motion between each successive splittings, conditioning
the walk to go through these splitting points.

\begin{figure}[ht]
  \begin{center}
    \includegraphics[width=0.65\textwidth]{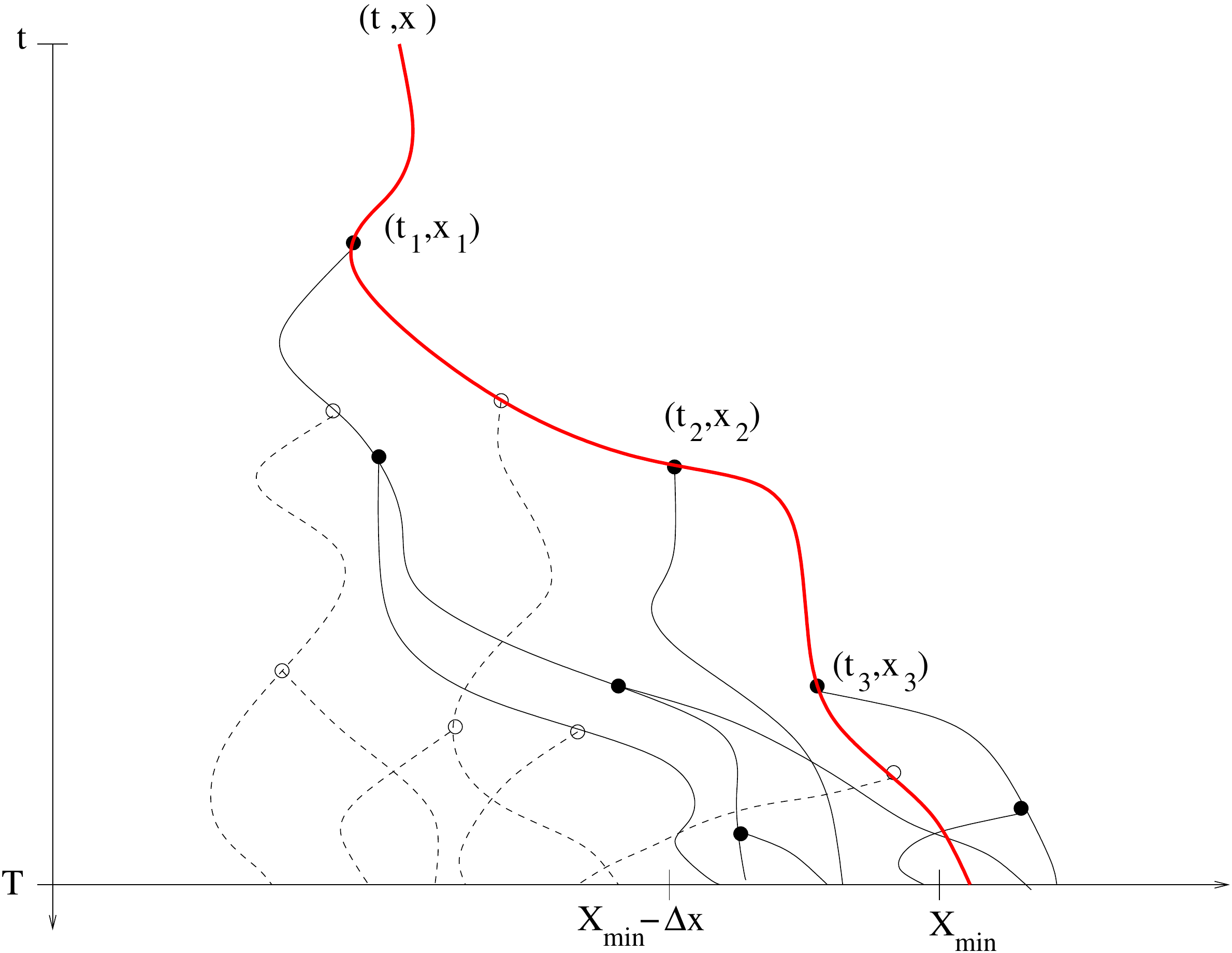}
  \end{center}
  \caption{\label{fig:scheme}
    {\small {Schematic evolution to time $T$ of a particle, starting
        at position $x$ at time $t$, through the branching Brownian motion
        conditioned to having at least one (``marked'') particle
        to the right of $\Xmin$ at $T$.}
      The horizontal axis carries the space variable, the vertical one the
      time, which flows from top to bottom.
      The trajectory of this marked particle is represented by the thick full line.
      Its offspring may a priori end up at any position. The branchings that generate
      particles at position not smaller than $\Xmin-\Delta x$ are denoted by black disks,
      while the ones that generate only particles with smaller positions are denoted
      by circles. The trajectories of the latter are
      represented by dashed lines.
    }
  }
\end{figure}

A realization produced with such an algorithm is sketched in Fig.~\ref{fig:scheme}.
A realization actually generated, in the discretized version of the
algorithm described in the next section, is displayed in Fig.~\ref{fig:event}.

%%%%%%%%%%%%%%%%%%%%%%%%%%%%%%%%%%%%%%%%%%%%%%%%%%%%%%%%%%%%%%%%%%%%%%

\subsection{\label{sec:discrete}Discrete model}

Although the algorithm described above can in principle be directly
implemented, in practice, it is quite challenging to achieve
a good accuracy. The reason is that a solution to the
continuous FKPP equation is needed, potentially at any point in space-time.
Such a solution can be obtained numerically only in
an approximate way, by discretizing the FKPP equation on a lattice.
Thus it is natural to start from the beginning
with a discrete model in space and time, in
such a way that the solution to the corresponding discretized FKPP equation
directly provides the exact (up to machine accuracy) probabilities of
the elementary processes involved.

We consider the following model discrete in space and time on a lattice
with respective spacing $\delta x$ and $\delta t$:
To evolve from $t$ to $t+\delta t$, a particle at $x$ may jump
to $x-\delta x$ with probability $p_l$ or to $x+\delta x$
with probability $p_r$ (these two processes define a random walk),
or may stay at the same position but split to two particles
with probability $p_b$.
The choice of the parameters of the model is constrained by
the unitarity relation $p_l+p_r+p_b=1$
which must be imposed, together of course with the restriction
of each of these elementary probabilities to the interval $[0,1]$.

The detailed implementation of this
model depends on the observable we want to measure.
We are first going to briefly show how to implement a full unbiased
evolution, and then describe the method to generate
ensembles of ``long-tail'' realizations
in which there is at least one particle 
in the interval $[\Xmin,+\infty[$ at the final time~$T$.

%%%%%%%%%%%%%%

\subsubsection{\label{sec:unbiased_algo}Generating unconditioned realizations}

At time $t$, the system is fully characterized by the set of
the numbers of particles $\{n(t,x)\}$ on all occupied sites.
The evolution of the system
between the times $t$ and $t+\delta t$ goes as follow:
Among the $n(t,x)$ particles on site $x$ at time $t$, there are 
$n_l(t,x),n_r(t,x)$ particles that
move on the site to the left
and to the right respectively, and $n_b(t,x)$ particles that
duplicate, with the sum rule $n_l+n_r+n_b=n$.
The numbers $(n_l,n_r,n_b)$ are distributed according to the multinomial
law
\be
\text{proba}(n_l,n_r,n_b=n-n_l-n_r)=
\frac{n!}{n_l!\, n_r!\, (n-n_l-n_r)!}\,
p_l^{n_l}p_r^{n_r}(1-p_l-p_r)^{n-n_l-n_r}.
\label{eq:p_unbiased}
\ee
On the sites on which $n$ is very large compared to unity,
the stochastic evolution can safely be
replaced by the mean-field deterministic evolution.

The numbers $\{n_l(t,x),n_r(t,x),n_b(t,x)\}$ are drawn
randomly according to the law~(\ref{eq:p_unbiased}) at each time step,
except on the sites on which $n(t,x)$ is very large, in which case
the numbers $n_l(t,x)$, $n_r(t,x)$ and $n_b(t,x)$ are just
proportional to $n(t,x)$, with respective coefficients $p_l$, $p_r$ and $p_b$
(``mean field'' approximation).
In any case, for each new iteration, the CPU time is proportional to
the number of occupied sites, which is a stochastic variable
the mean of which grows linearly with the evolution
time. Hence the overall complexity grows quadratically with the final time $T$.
The memory occupancy depends on the observable. It is linear in~$T$
for the observables of interest in this work, for which we just need to
record the particle content at the current time.

%%%%%%%%%%%%%%%%%%%%%%5

\subsubsection{\label{sec:algo}Conditioned ensembles}

Long-tail events are (arbitrarily) rare, so it is not possible to build
a large statistical ensemble of such realizations using the full Monte Carlo
we have just described.
It is however possible to write an exact Monte Carlo that generates
{\it only} events which have at least one
particle with position not smaller than $\Xmin$ at the final time~$T$.
We can generate complete realizations, that is to say, keep
all the particles until the final time; We can also evolve only the
particles that end up at positions not smaller than say $\Xmin-\Delta x$ at time~$T$,
which is actually enough if the purpose is to study the tip of the particle
distribution at the final time. We shall now describe the two versions
of the algorithm in greater detail.

\paragraph{Complete realizations.}

The initial particle is marked and conditioned to arrive at some
position~$x\geq \Xmin$. This ensures that in the realization, there
is one or more particle(s) in that region.
Its evolution is conditioned to having at least one
particle in its descendence
in that spatial region. As for the first step,
which brings the marked particle from time~$t$ to time $t+\delta t$,
this conditioning
is performed through the replacement of
the probabilities $p_l$, $p_r$ of the elementary jump processes
by
\be
\begin{split}
  &p_l^\prime(T-t,\Xmin-x)\equiv
  p_l\times \frac{u(T-t-\delta t,\Xmin-x+\delta x)}{u(T-t,\Xmin-x)}\\
  &p_r^\prime(T-t,\Xmin-x)\equiv
  p_r\times \frac{u(T-t-\delta t,\Xmin-x-\delta x)}{u(T-t,\Xmin-x)},
\end{split}
\label{eq:biased_p}
\ee
and the branching probability by
\be
 p_b' \equiv 1-p_l'-p_r'.
\ee
In these expressions, $u(\tau,\xi)$ is the probability of having
at least one particle with position not smaller than~$\xi$
at time $\tau$ starting with a single particle at position $\xi=0$ at time $\tau=0$.
It solves an equation in the universality class of
the FKPP equation that is straightforward to derive from the 
definition of the model,
\be
u(\tau+\delta t,\xi)=p_r\, u(\tau,\xi-\delta x)+p_l\, u(\tau,\xi+\delta x)
+p_b\,u(\tau,\xi)[2-u(\tau,\xi)],
\label{eq:dFKPP}
\ee
with the initial condition
\be
u(0,\xi)=1\quad\text{for $\xi\leq 0$},\quad
u(0,\xi)=0\quad\text{for $\xi>0$}.
\ee

Using the probabilities $p'_l$, $p'_r$, $p_b'$,
one chooses among the three possible outcomes of the
evolution between times $t$ and $t+\delta t$:
The particle moves left,
or moves right, or branches.
As before for the BBM, there are two main cases to distinguish:
\begin{itemize}
\item If no branching occurs, the position of the marked particle is just updated
from $x$ to $x\mp\delta x$,
according to whether it jumps left or right. The next evolution step,
to the time $t+2\delta t$,
uses the biased probabilities $p'_l$, $p'_r$ and $p_b'$ evaluated at the point
$(T-t-\delta t,\Xmin-x\pm\delta x)$.

\item
  If a branching occurs, then there is a second particle at position $x$ at time $t+\delta t$,
in addition to the marked one. This new particle is evolved with
the full unbiased algorithm described in the previous section~\ref{sec:unbiased_algo}
until the final time $T$, independently of the marked particle.
The marked particle instead keeps evolving according to the biased
probabilities $p'_l$, $p'_r$ and $p_b'$, evaluated at
$(T-t-\delta t,\Xmin-x)$.
\end{itemize}

\paragraph{Keeping only the particles close to the tip.}

For our particular purpose in this paper, and actually for all studies which
would focus only on the region near the lead particle, it
is enough to keep track of the particles which have positions not smaller than
$\Xmin-\Delta x$ at the final time $T$. Hence when a branching of the
marked particle occurs, it is enough to keep the offspring only if it possesses at least
one descendant with position not smaller than
$\Xmin-\Delta x$ at the final time $T$.

The jumps of the marked particle keep the same probabilities as given
in Eq.~(\ref{eq:biased_p}).
The branching process of the marked particle is replaced by two processes:
A branching may still occur, but with the probability
\begin{multline}
p_b'(T-t,\Xmin-x)\equiv p_b\times
u(T-t-\delta t,\Xmin-x)\\
\times
\frac{2u(T-t-\delta t,\Xmin-\Delta x-x)-u(T-t-\delta t,\Xmin-x)}
{u(T-t,\Xmin-x)},
\end{multline}
which takes into account the new conditioning.
But the marked particle may also just remain on its site $x$ without branching.
This may occur with the probability
\be
p_s'\equiv 1-p'_l-p'_r-p_b'.
\ee

When a branching of the marked particle takes place, while the
evolution of the latter is just an iteration until the final time~$T$
of the processes we have
just described,
the further evolution of the second particle is conditioned in such a way that it
has at least
one descendent which arrives at a position not smaller than $\Xmin-\Delta x$.
Hence from the time $t+\delta t$ to $t+2\delta t$,
it evolves according to the following processes: It may move left
or right with respective probabilities
\be
p_l''\equiv p'_l(T-t-\delta t,\Xmin-\Delta x-x)
\quad\text{and}\quad
p_r''\equiv p'_r(T-t-\delta t,\Xmin-\Delta x-x),
\ee
or branch to two particles (that both get further evolved)
with probability
\be
p_b''(T-t-\delta t,X-x)\equiv p_b\times
\frac{\left[u(T-t-2\delta t,\Xmin-\Delta x-x)\right]^2}{u(T-t-\delta t,\Xmin-\Delta x-x)},
\ee
or else, do nothing, with probability
\be
p_s''\equiv 1-p''_l-p''_r-p_b''.
\ee

\begin{figure}
  \begin{center}
    \includegraphics[width=.9\textwidth]{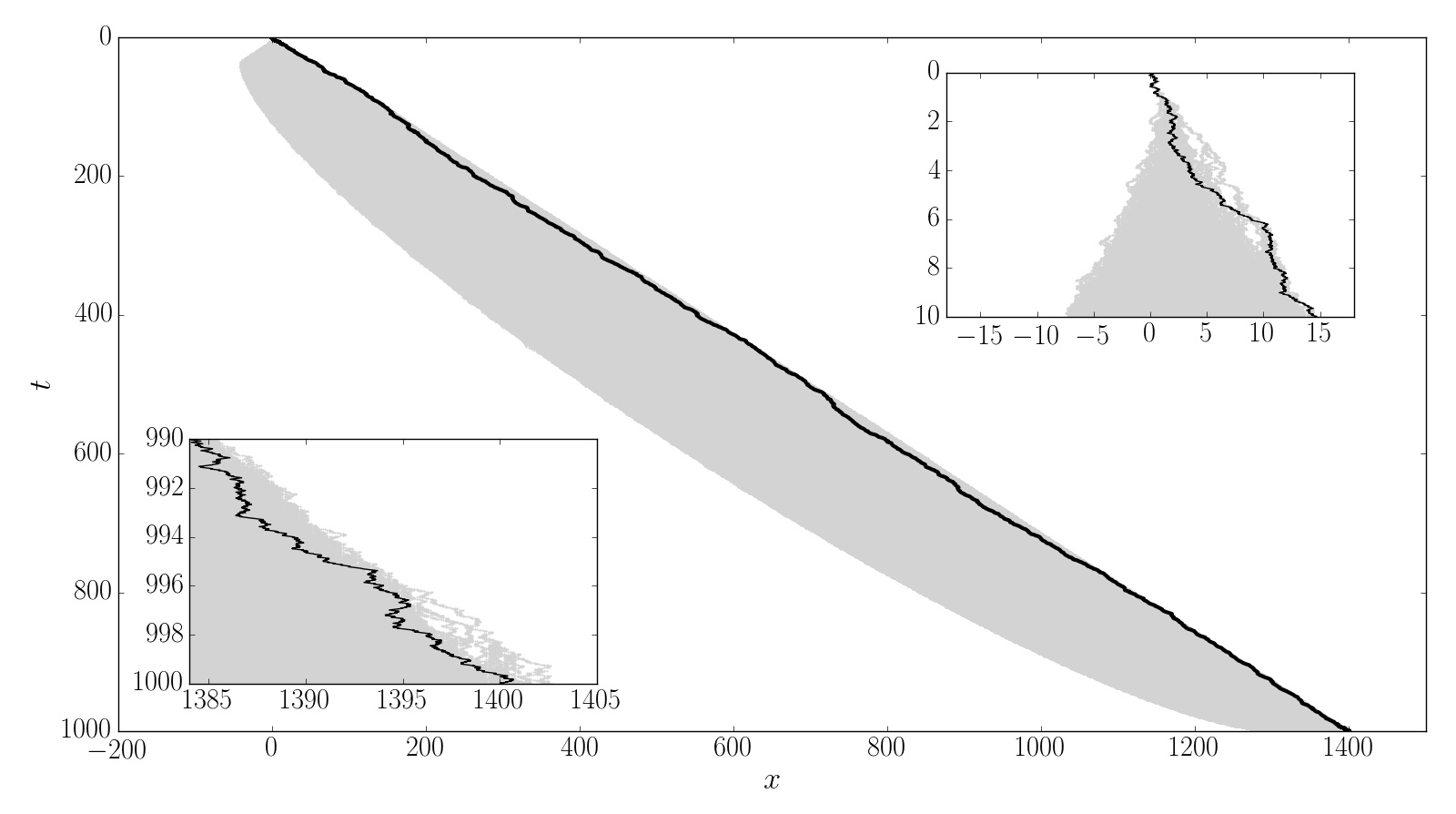}
  \end{center}
  \caption{
    \label{fig:event}
    {\small
      Realization of the BRW  defined in Sec.~\ref{sec:discrete} and~\ref{sec:analytic}
      evolved to $T=1000$ using the algorithm described in Sec.~\ref{sec:algo}.
      Time flows from top to bottom. $\Xmin$ is set to 1400, and one keeps only
      the particles that end up at a position not smaller than $1300$.
      The grey zone represents
      the set of the space-time trajectories
      of all the particles that are evolved by the algorithm.
      The black continuous line
      represents the worldline of the marked particle,
      that is conditioned to end up
      at a position greater
      or equal to $\Xmin$ at time~$T$. {\it Insets:} Zoom on the initial times
      $t\in[0,10]$ (upper right)
      and final times $t\in[990,1000]$ (lower left).
    }
  }
\end{figure}

By iterating these processes until the final time~$T$ is reached,
one only evolves the particles that end up
at time $T$ with a position greater or equal to $\Xmin-\Delta x$,
including a marked particle which ends up at
a position greater or equal to $\Xmin$.
This procedure leads to realizations of particle sets
in the range of positions
$[\Xmin-\Delta x,+\infty[$ at time~$T$ distributed according to the exact
    probabilities.
    One particular realization generated by our code is shown in Fig.~\ref{fig:event}.

The main limitation of this algorithm is the size of the memory needed to store
the solution to Eq.~(\ref{eq:dFKPP})
(which may be traded for CPU time by partially recomputing~$u$
when needed), which grows linearly with $T$.

%%%%%%%%%%%%%%%%%%%%%%%%%%%%%%%%%%%%%%%%%%%%%%%%%%%%%%%%%%%%%%%%%%%%%%

\section{\label{sec:results}
  Probability distribution of particle numbers near the tip}

We consider a BRW evolved from time~0 to time $T$.
We denote by $X$ the position of the rightmost particle at $T$ in a given
realization.
The observable we are interested in is
the probability distribution $p_n(\Delta x)$ of the number $n$ of particles in
the interval $[X-\Delta x,X]$. The position
$X$ can be let free, or can be constrained to
be at least as large as some minimal value $\Xmin$.

\subsection{\label{sec:analytic}Analytical expressions for a few observables}

The analytical expressions for the typical and mean particle numbers
involve the values of a few parameters
that characterize the shape and velocity of the traveling wave\footnote{
  The properties of solutions to the FKPP equation~(\ref{eq:FKPP_Q})
  at large times were first worked out in Ref.~\cite{Bramson}.
  These properties are shared by a wide class of models,
  see e.g. Ref.~\cite{VANSAARLOOS200329}. More detailed studies have appeared
  recently, see e.g.~\cite{Graham_2019,Berestycki_2018}.
  }
which is known to solve asymptotically Eq.~(\ref{eq:dFKPP}).
We start by computing these parameters for the model we have implemented,
and then, we briefly review
the available analytical results.

\subsubsection{Properties of the asymptotic solutions to the FKPP equation}

The method to compute the parameters characterizing the asymptotic solutions
to equations in the universality class of the FKPP equation
is standard. We will not expose it in detail: Instead, we shall refer the reader to
a review such as the one in Ref.~\cite{VANSAARLOOS200329}.

One looks for solutions to the equation obtained by linearizing
equation~(\ref{eq:dFKPP}) near the fixed point $u=0$,
using the {\it Ansatz} $e^{-\gamma[x-v(\gamma)t]}$. The
velocity $v(\gamma)$ of the partial wave $e^{-\gamma x}$ reads
\be
v(\gamma)
=\begin{cases}
\frac{\gamma}{2}+\frac{1}{\gamma} &\text{for the BBM}\\
\frac{1}{\gamma\delta t}\ln
\left(p_r\,
  e^{\gamma\delta x}+
  p_l\,e^{-\gamma\delta x}+2p_b
  \right)& \text{for the BRW}.
\end{cases}
\ee
While the linearized equation results in a (discretized) diffusion
and an exponential growth of~$u$ with time due to the branchings, the main
effect of the nonlinearity present in Eq.~(\ref{eq:dFKPP})
is to make sure that $u$ remains less than~1
by compensating for its exponential growth with time
in the spatial regions in which $u\sim {\cal O}(1)$. The result
of the joint action of the linear and nonlinear terms in Eq.~(\ref{eq:dFKPP})
is the formation of a so-called ``traveling wave'', namely a moving front
smoothly connecting $u=1$ at $x=-\infty$ to $u=0$ at $x=+\infty$.
The transition region between these asymptotics\footnote{The position
  of the front is ambiguous
  and needs to be defined. One may choose it to coincide with the value
  of $x$ for which $u(t,x=m_t)=\frac12$, for example.
}
turns out to be located at the time-dependent spatial coordinate
\be
m_t=v(\gamma_0) t-\frac{3}{2\gamma_0}\ln t+\text{const},
\label{eq:mt}
\ee
up to terms that vanish in the limit $t\rightarrow+\infty$.
In this formula, the additive constant
depends on the very definition of the front position, the two other terms
are universal. The constant
$\gamma_0$ is the value of $\gamma$ which minimizes $v(\gamma)$.
The wave front has the shape
\be
u(t,x)\simeq\text{const}\times(x-m_t)\,
e^{-\gamma_0(x-m_t)}\exp\left(
-\frac{(x-m_t)^2}{2\gamma_0 v''(\gamma_0)t}
\right)
\label{eq:shape_traveling}
\ee
in the parametric region
$1\ll x-m_t\lesssim \sqrt{\gamma_0 v''(\gamma_0)t}$ and for large
times $t\gg 1$.

In practice, we set the parameters of the model as follows:
\be
p_r=p_l=\frac12(1-\delta t),\
p_b=\delta t,
\quad\text{with}\quad
\delta t=0.01,\ \delta x=0.1.
\label{eq:p}
\ee
The minimization condition for the velocity function,
$v'(\gamma)|_{\gamma=\gamma_0}=0$, has the following solutions,
in the two cases of interest in this paper:
\be
\gamma_0=\begin{cases} \sqrt{2}&\text{for the BBM}\\
1.4319525\cdots&\text{for the BRW}.
\end{cases}
\label{eq:pardiscrete1}
\ee
This constant sets the decay rate of the wave front at very
large $t$ and for $x-m_t\gg 1$.
The other model-dependent constants entering
Eqs.~(\ref{eq:shape_traveling}),(\ref{eq:mt}) read
\be
v(\gamma_0)=\begin{cases}
\sqrt{2}\\
1.3943622\cdots,
\end{cases}
\gamma_0\, v''(\gamma_0)=\begin{cases}
1 & \text{for the BBM}\\
0.96095291\cdots & \text{for the BRW.}
\end{cases}
\label{eq:pardiscrete2}
\ee

We may observe
that with our choice~(\ref{eq:p}), the particular BRW model
we implement can indeed be seen
as a discretization of the BBM with diffusion constant~$\frac12$ and
replacement rate~1. As expected, the constants
in Eqs.~(\ref{eq:pardiscrete1}),(\ref{eq:pardiscrete2}) differ by
a few percents between the BBM and the BRW,
as an effect of the discretization in~$t$ and~$x$.
Note that it is essential to take time steps of size $\delta t$
small compared to the inverse branching rate, although
such choices make it more difficult to reach large times in the
numerical calculation.

%%%%%%%%%%%%%%%

\subsubsection{Characterizing analytically the particle number at the tip}

We now review a few properties of the distribution of the particle number~$n$
in the interval $[X-\Delta x,X]$~\cite{Brunet2011,Mueller:2019ror}.

\paragraph{Case in which $X$ is unconstrained.}

In the case in which the realizations are not conditioned to the final
position of the rightmost particle, the measure of the
particle density as seen from the tip was shown to be
a decorated Poisson process~\cite{Brunet2011} (see also
\cite{arguin2012,aikedon2013,Arguin2013}).
We shall denote by $p_n^{(0)}(\Delta x)$ the distribution of the number of particles
in an interval $\Delta x$ from the tip, and $\bar n^{(0)}(\Delta x)$
its expectation value.

In the region $1\ll\Delta x\lesssim \sqrt{\gamma_0v''(\gamma_0)T}$,
the expectation value of~$n$ scales like\footnote{
  This can easily be understood in a picture in which the branching random
  walk is replaced by a deterministic evolution with an absorptive boundary
  at its tip modeling discreteness of the original stochastic process,
  see e.g. Ref.~\cite{Mueller:2014gpa}.
  }
\be
\bar n^{(0)}(\Delta x)=\text{const}
\times\Delta x\,e^{\gamma_0\Delta x} e^{-{\Delta x^2}/(2\gamma_0v''(\gamma_0)T)}.
\label{eq:n0bar}
\ee
(The $T$-dependence in the r.h.s., which is a finite-$T$ correction going away
for $T=+\infty$, is
kept implicit in the l.h.s).

We may derive the shape of the distribution $p_n^{(0)}(\Delta x)$
from a simple picture.
Let us define
\be
Z_T\equiv\sum_{i=1}^{N(T)} [v(\gamma_0)T-x_i(T)] e^{\gamma_0 [x_i(T)-v(\gamma_0)T]},
\ee
where $\{x_i(T),1\leq i\leq N(T)\}$ is the set of the positions of the
$N(T)$ particles at time~$T$ in the particular realization we are considering.
According to the Lalley and Sellke theorem~\cite{lalley1987},
for $T$ large,\footnote{%
  Strictly speaking, the theorem holds in the limit $T\rightarrow +\infty$.}
the $T$-dependent random variable\footnote{%
  Some statistical properties of $Z_T$ were conjectured
  in Ref.~\cite{Mueller:2014gpa}
  and proven rigorously in Ref.~\cite{maillard2019}.
  }
$Z_T$ converges to a $T$-independent but realization-dependent
positive number~$Z$, with
probability~1. Furthermore,
there is a frame the origin of which is, at time $T$, at position
\be
X_T=m_T+\frac{1}{\gamma_0}\ln Z+\text{const},
\label{eq:LSframe}
\ee
with respect
to which the probability density of $X$ obeys
the Gumbel distribution\footnote{
Note that Gumbel distributions~\cite{AIHP_1935__5_2_115_0}
appear generically in the statistics of extremes
for a wide class of problems; See for example Ref.~\cite{Bouchaud_1997}
or the very recent review
in Ref.~\cite{Majumdar_2020}.
}
\be
\gamma_0\exp\left(-\gamma_0(X-X_T)-e^{-\gamma_0(X-X_T)}
  \right).
\label{eq:gumbel}
\ee
The additive constant in Eq.~(\ref{eq:LSframe}) is independent of the realization.

In this frame, the number of particles within the interval $[-\xi,0]$ (with $\xi>0$)
in the realization grows like $\bar n^{(0)}(\xi)$,
up to negligible fluctuations when $\xi$ is large enough.
Hence the number of particles at a distance $\Delta x$ from the tip at position $X$
is proportional to $\bar n^{(0)}(\Delta x-X+X_T)$.
From the distribution~(\ref{eq:gumbel}) of $X-X_T$,
we get, in the large-$T$ and large-$\Delta x$ limit,\footnote{
Multiplicity distributions that depend on the
parameters only through the mean multiplicity, as in Eq.~(\ref{eq:pn0}),
are frequent in particle physics: This phenomenon is known as
the ``Koba-Nielsen-Olesen (KNO) scaling''. It was found to hold in many
experimental measurements of distributions of numbers of
particles produced in scattering processes~\cite{KOBA1972317}.}
\be
\text{proba}\left(\delta\equiv\ln\frac{n}
     {\bar n^{(0)}(\Delta x)}\right)
     =\exp\left({\delta-e^\delta}\right)
     \implies
     \text{proba}(n)=\frac{1}{\bar n^{(0)}(\Delta x)}
     \exp\left(-\frac{n}{\bar n^{(0)}(\Delta x)}\right).
\label{eq:pn0}
\ee

In the picture we have used here, the fluctuations of the multiplicity $n$
just reflect the fluctuations of the
position of the lead particle in the Lalley-Sellke frame defined in
Eq.~(\ref{eq:LSframe}).

\paragraph{Case in which $X-m_T$ is set large.}

A few properties of the particle number distribution in the interval
$[X-\Delta x,X]$ have been
either computed or conjectured~\cite{Mueller:2019ror}
in the joint limits $T\rightarrow +\infty$, $\Delta x\gg 1$, in the case
in which the interval is included in the
region in which the solution of the FKPP equation
has reached its asymptotic shape $\propto (x-m_T)\,e^{-\gamma_0(x-m_T)}$,
namely in which the Gaussian factor that exhibits an explicit $T$-dependence
is very close to~1.
This region is often called the ``scaling region'':
It extends over a
distance of the order of $\sqrt{\gamma_0\,v''(\gamma_0)\,T}$
to the right of the mean position $m_T$ of the rightmost particle.

As for the expectation value of~$n$, 
\be
\bar n(\Delta x)=\text{const}\times e^{\gamma_0\Delta x},
\label{eq:mean_conditioned}
\ee
up to finite-$T$ and finite-$\Delta x$ corrections
that should vanish in the asymptotic limits.
As for the typical value, 
the following conjecture was proposed:
\be
n_{\text{typical}}=\text{const}\times\bar n(\Delta x)
\times e^{-\zeta\Delta x^{2/3}},
\label{eq:n_typical}
\ee
again up to finite-$T$ corrections, and up to
finite-$\Delta x$ corrections that might take the
form of a linear prefactor.

The method to arrive at these results was based on the observation,
derived from the discussion in Ref.~\cite{Brunet2011},
that a generating function of the particle number probabilities
near the tip,
\be
G_{\Delta x}(\lambda)=\sum_{n=1}^\infty \lambda^n p_n(\Delta x),
\ee
can be deduced from the solution
to the FKPP equation with a peculiar $\lambda$- and $\Delta x$-dependent
initial condition.
The infinite-$T$ and large-$\Delta x$ behavior of the solution were then
analyzed
in two limits of the parameter $\lambda$ of the generating
function: $\lambda\rightarrow 1$ and $\lambda\ll 1$~\cite{Mueller:2019ror}.
The former gives access to the mean particle number,
while in the latter limit, a scaling
form for $G$ was found, from which one could conjecture the behavior of
the typical particle number~\cite{Mueller:2019ror}.

%%%%%%%%%%%%%%%%%%%%%%%%%%%%%%

\subsubsection{\label{sec:typical}
  Estimator of the typical value of the particle number}

We need to specify what we mean by ``typical value'' of $n$ quantitatively.

With the reasonable assumption (well-verified numerically, see below)
that $p_n(\Delta x)$ as a function of~$n$ has a unique maximum,
we may define it to be the mode of the distribution, that is to say,
the value $n_{\text{typical}}^{\text{max}}$ of
the particle number $n$ that maximizes $p_n(\Delta x)$. This number is
obviously defined by the equation
\be
\left.\frac{d}{dn}\right|_{n=n_{\text{typical}}^{\text{max}}}p_n(\Delta x)=0
\ee
when $n$ is large enough so that $p_n(\Delta x)$ can be considered a continuous
(differentiable) function of $n$. But this is awkward
when one works with statistical samples of moderate size: Therefore, we will not use
this prescription in practice.

We have tried instead the median
$n_{\text{typical}}^{\text{median}}$, defined by the equations
\be
\text{proba}\left(n\leq n_{\text{typical}}^{\text{median}}\right)\geq\frac12
\quad\text{and}\quad
\text{proba}\left(n\geq n_{\text{typical}}^{\text{median}}\right)\geq\frac12.
\label{eq:median}
\ee
Alternatively,
we have used the expectation value of $\ln n$:
\be
n_{\text{typical}}^{\text{log}}\equiv \exp\left({\langle\ln n\rangle}\right).
\label{eq:explog}
\ee
The latter is a good estimator of the typical value of $n$
for distributions which are symmetric in a logarithmic
scale for $n$. Empirically, this turns out to be the
case for the distribution of $n$ when
the tip is conditioned to be far from its typical or mean position.

%%%%%%%%%%%%%%%%%%%%%%%%%%%%%%%%%

\subsection{Numerical results}

We now use the code we have developed in order to test numerically the
formulae in Sec.~\ref{sec:analytic} in the case of the discrete model
described in Sec.~\ref{sec:discrete}.

The calculations/conjectures we want to test are actually valid
for $T\rightarrow +\infty$, which, needless to say, can never be achieved
numerically. What we can do instead is to test how the asymptotics is
approached by performing calculations for different increasing values
of the final evolution time $T$.

For each choice of the parameters $\Xmin$, $\Delta x$ and $T$,
we need ensembles of a few thousands of realizations.
With our algorithm and the state-of-the-art computer technology,
the evolution time $T$ that may be reached within a 
few thousands hours of computer time is of the order of $10^4$.

For given $\Xmin$ and $T$, we actually measure observables for the
different values of $\Delta x$ on the same samples of realizations.
Hence these measurements are correlated.

\subsubsection{Unconstrained realizations}

We first let the position~$X$ of the rightmost particle unconstrained.
We shall nevertheless use the algorithm described above but
with $\Xmin$ set to a value less than $m_T$, chosen in such a way that
the probability to have no particle to the right of $\Xmin$ be negligible.
The biasing has no effect on the observable since, with this procedure,
$X$ is effectively unconstrained,
but enables one to reduce
the complexity of the numerical calculation.

\paragraph{Probability density.}

\begin{figure}[ht]
  \begin{center}
    \begin{tabular}{cc}
    \includegraphics[width=.49\textwidth]{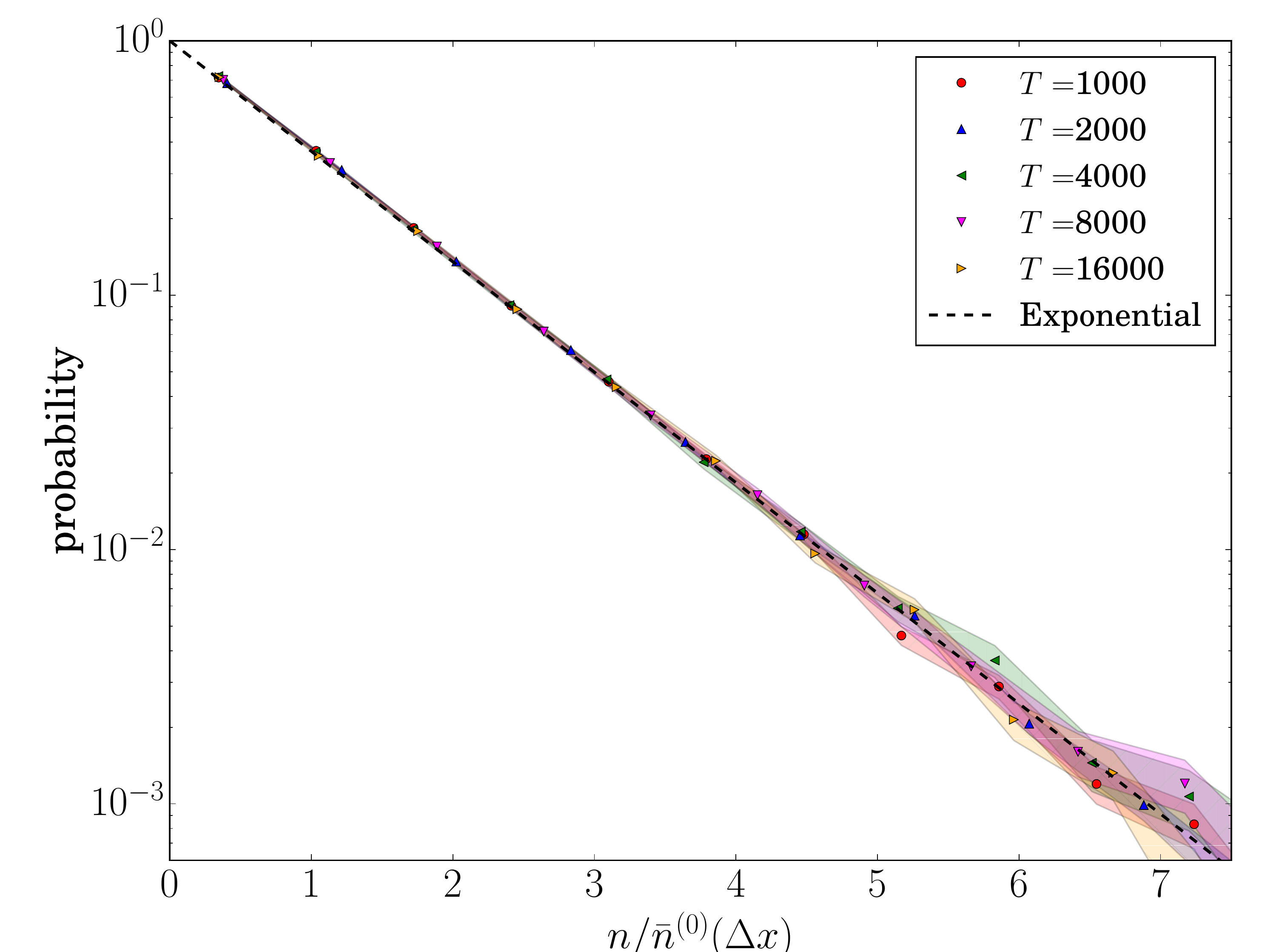}&
    \includegraphics[width=.49\textwidth]{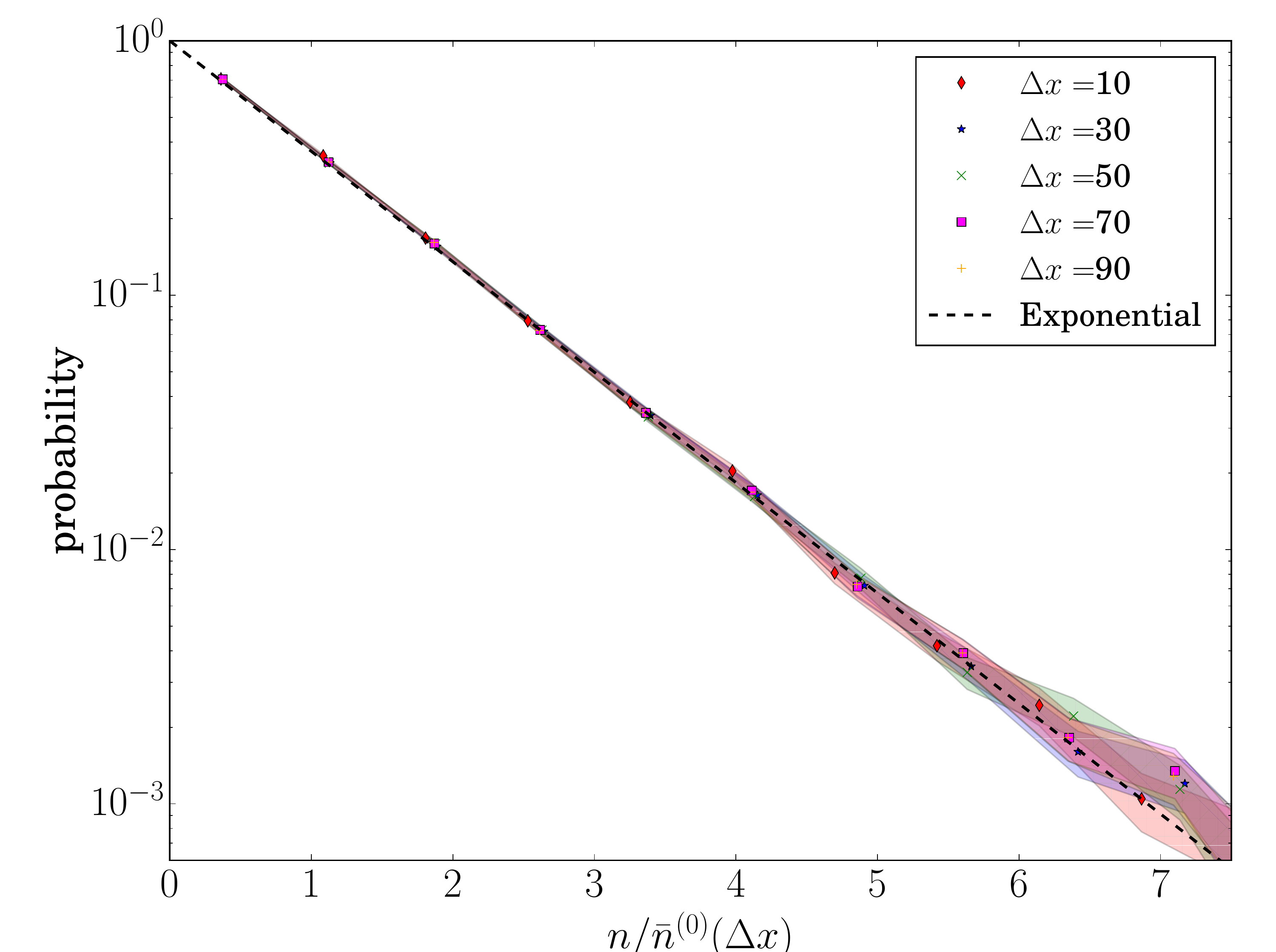}\\
    (a)&(b)
    \end{tabular}
  \end{center}
  \caption{\label{fig:unbiased_proba_exp}
    {\small
      Probability distribution of the number $n$ of particles
      in an interval of size $\Delta x$ to the left of the
      lead particle, at time $T$. The particle number $n$ is normalized by
      its expectation value $\bar n^{(0)}(\Delta x)$, computed
      numerically using the available data.
      (a) $\Delta x=30$ and different values of 
      the evolution time $T$.
      (b) $T=8000$ and different interval sizes $\Delta x$.
      The dashed line represents the exponential
      $e^{-n/\bar n^{(0)}(\Delta x)}/\bar n^{(0)}(\Delta x)$.
    }
    }
\end{figure}

The distribution $p_n^{(0)}(\Delta x)$ of the number~$n$
of particles in the interval $[X-\Delta x,X]$
is shown in Fig.~\ref{fig:unbiased_proba_exp}.
The points represent the probability of observing a given value of $n$ in bins
of fixed size on the scale $n/\bar n^{(0)}(\Delta x)$. In this plot,
the expectation value of the particle number is evaluated on the statistical
ensemble. The error band is computed assuming
Gaussian uncertainties in each bin.

We see that all the data points, either for fixed $\Delta x=30$ and
different values of $T$ (Fig.~\ref{fig:unbiased_proba_exp}a) or
for fixed $T=8000$ and different values of $\Delta x$
(Fig.~\ref{fig:unbiased_proba_exp}b), fall on the same exponential curve,
and are thus perfectly consistent with Eq.~(\ref{eq:pn0}).

\begin{figure}[ht]
  \begin{center}
    \begin{tabular}{cc}
    \includegraphics[width=.49\textwidth]{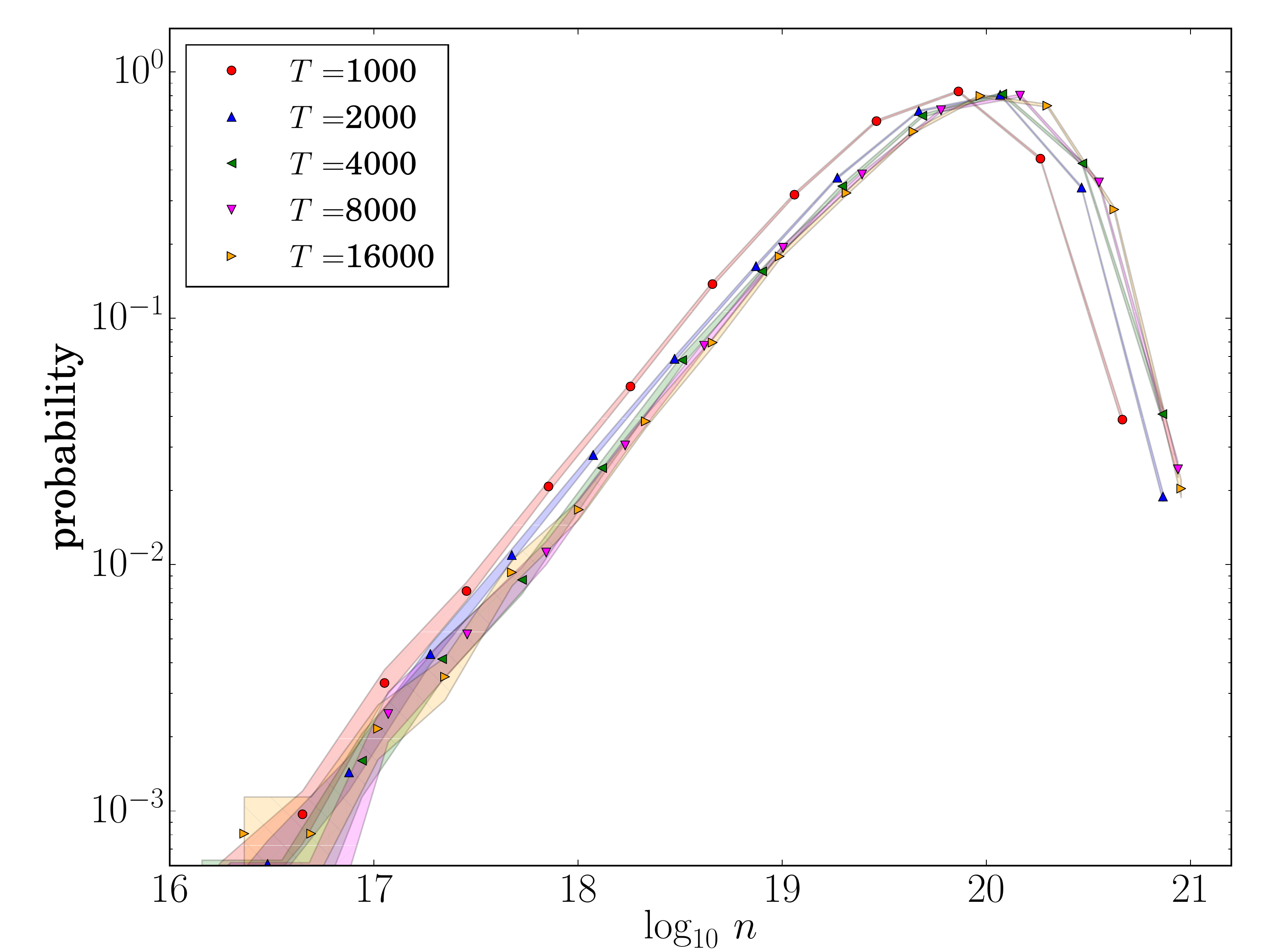}&
    \includegraphics[width=.49\textwidth]{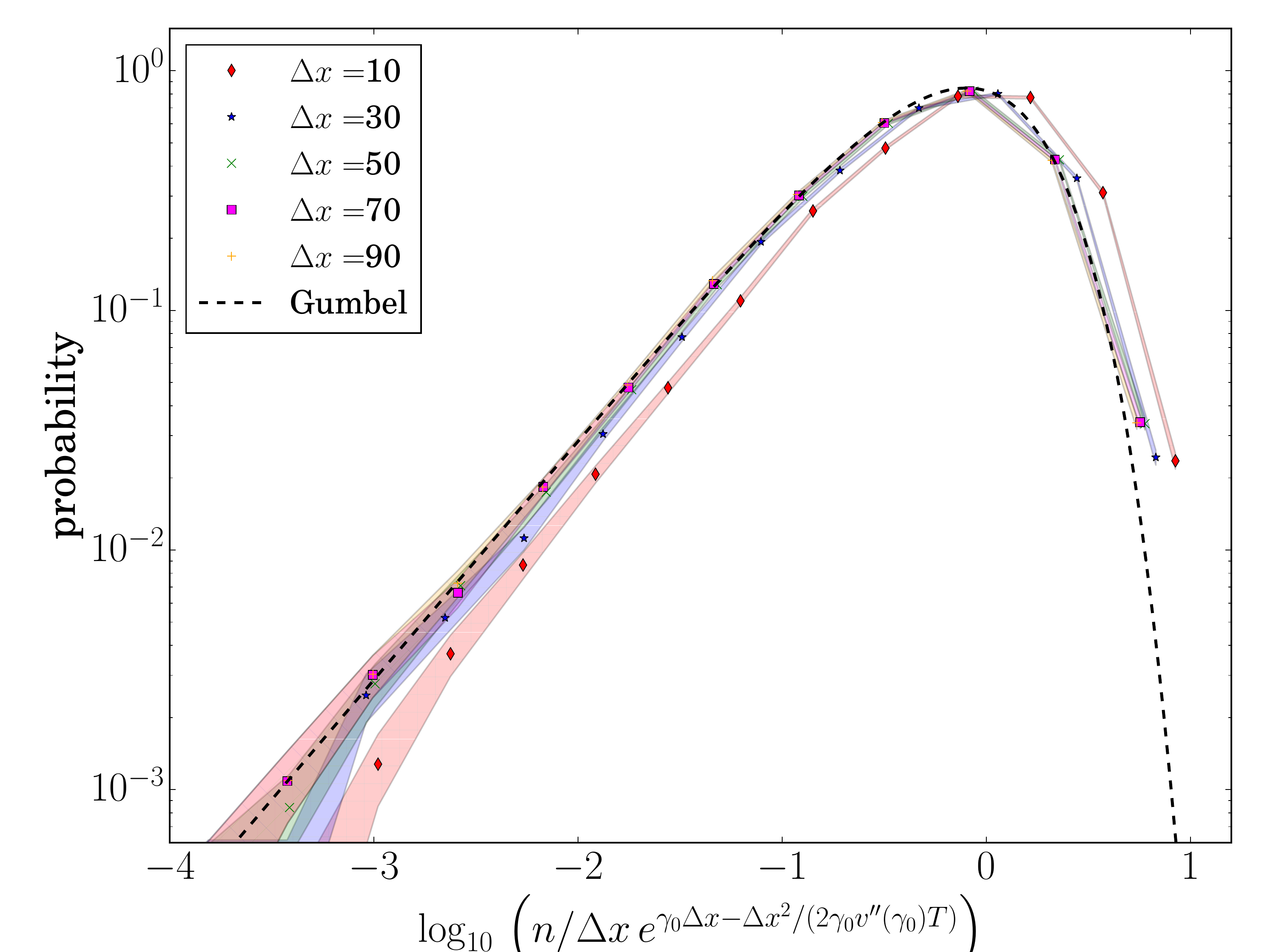}\\
    (a)&(b)
    \end{tabular}
  \end{center}
  \caption{\label{fig:unbiased_proba}
    {\small
      Probability distribution of the logarithm of the particle multiplicity
      in an interval of size $\Delta x$ to the left of the
      lead particle, at time $T$.
      (a) $\Delta x=30$ and different values of 
      the evolution time $T$.
      (b) $T=8000$ and different interval sizes $\Delta x$; The $x$-axis
      is translated by a term proportional to the expected mean particle number.
      The dashed line represents the theoretical probability~(\ref{eq:gumbel2})
      obtained in the simple picture explained in the text,
      with one free numerical constant of order~1 set by hand.
    }
    }
\end{figure}

For future comparison with the case in which $\Xmin$ is set large,
it is useful to also
plot the distribution of the logarithm of the particle number~$n$.
This is displayed in Fig.~\ref{fig:unbiased_proba}.
We see that the data is independent
of $\Delta x$ for large $\Delta x$, and is consistent with the
analytic form given in Eq.~(\ref{eq:pn0}).
Indeed, the curve representing the distribution
\be
\frac{\ln 10}{c}10^{\delta}\exp\left(-\frac{1}{c} 10^{\delta}\right)
\quad\text{with}\quad
\delta\equiv
\log_{10}\frac{n}{\Delta x\, e^{\gamma_0\Delta x-\Delta x^2/(2\gamma_0 v''(\gamma_0)T)}}
\label{eq:gumbel2}
\ee
fits very well the numerical data 
when the constant $c$ is set to 0.8.

As a consequence of the shape of the multiplicity distribution, the
typical and the mean values of $n$ are both proportional to
$\Delta x\, e^{\gamma_0\Delta x}$, supplemented by the expected
finite-$T$ correction factor
$e^{-\Delta x^2/(2\gamma_0 v''(\gamma_0)T)}$.

\paragraph{Mean and typical values.}

\begin{figure}[ht]
  \begin{center}
    \includegraphics[width=\textwidth]{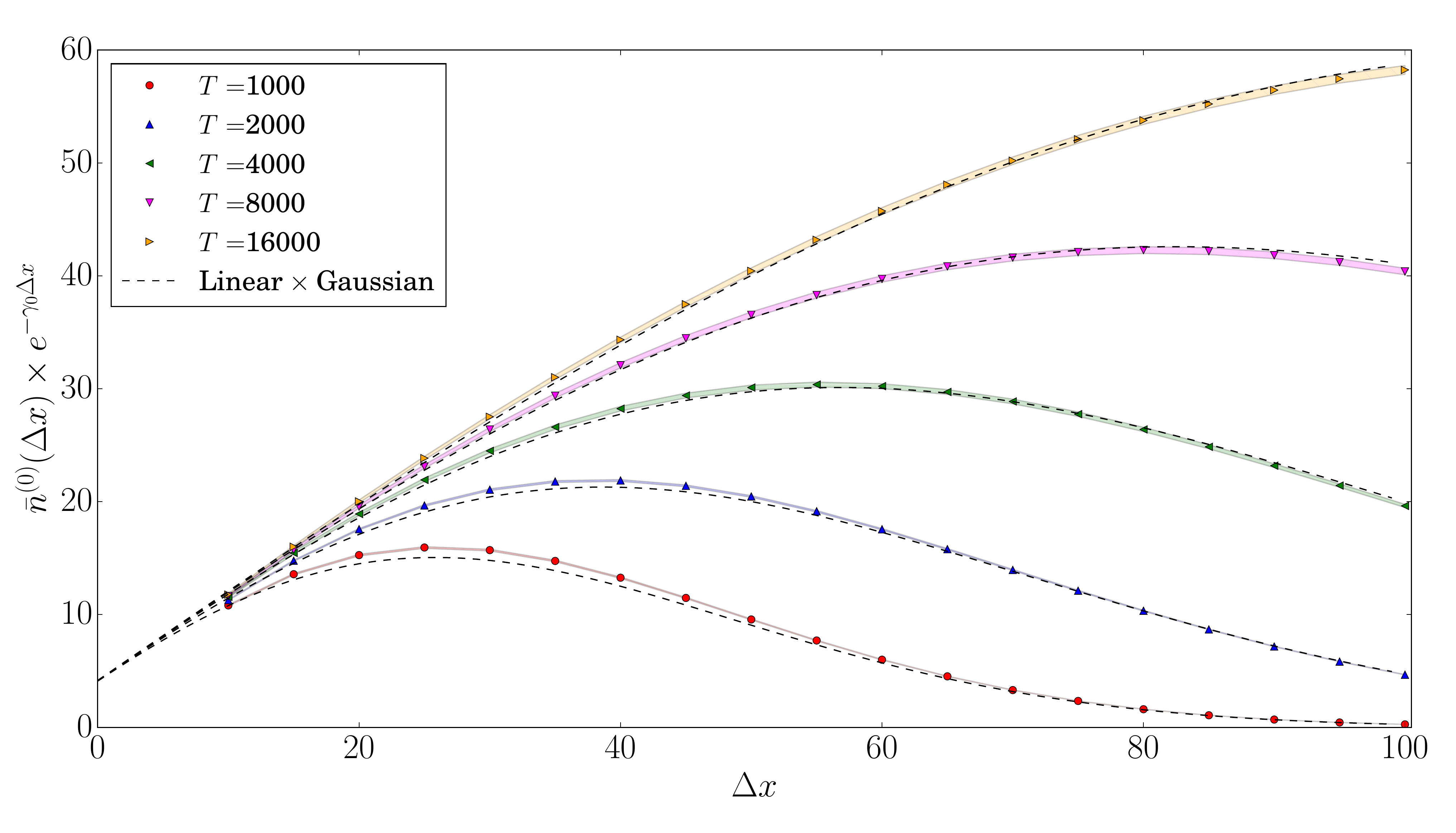}
  \end{center}
  \caption{\label{fig:unbiased_mean}
    {\small
    Mean value of the particle number as a function
    of $\Delta x$, for different values of the evolution
    time $T$.
    The dotted curves represent the expected shape in the limit
    of large $T$ and $\Delta x\gg 1$, up to two free constants,
    see Eq.~(\ref{eq:linear+Gaussian}).
    }
  }
\end{figure}

The mean value of the particle number, rescaled by $e^{-\gamma_0\Delta x}$, is shown in
Fig.~\ref{fig:unbiased_mean}  as a function of~$\Delta x$,
for different values of~$T$.
The error band is evaluated using the jackknife method.

To the data, we superimpose the curves
\be
\bar n^{(0)}(\Delta x)\times e^{-\gamma_0\Delta x}=\bar c(\Delta x+\bar a)
\exp\left(-\frac{(\Delta x+\bar a)^2}{2\gamma_0 v''(\gamma_0)T}\right)
\label{eq:linear+Gaussian}
\ee
where the normalization $\bar c$ is not computable and hence has to be
fitted, and $\bar a$ is a fudge term meant to effectively
take into account some subleading corrections. We determine these constants
using the data for $T=8000$. The best fit is obtained for
the values $\bar c\simeq 0.80$ and $\bar a\simeq 5.2$. Note that the value
of $\bar c$ fitted here is identical to the value of $c$ chosen
to describe the data for the probability distribution $p_n^{(0)}(\Delta x)$
(see Eq.~(\ref{eq:gumbel2}) and Fig.~\ref{fig:unbiased_proba}).

We see in Fig.~\ref{fig:unbiased_mean}
that the numerical data is consistent with this formula.\footnote{
  We could easily get a much better fit by refitting the parameters
  for each value of~$T$, and/or by introducing a third parameter instead
  of $\bar a$ in the
  shift of $x$ in the Gaussian factor. But our goal here is just to show
  consistency with a theoretical formula expected to hold for asymptotically
  large $T$ and $\Delta x$, limit
  in which the parameter $\bar a$ should be irrelevant.
  }
In particular, for the largest value of $T$, the linear factor
in Eq.~(\ref{eq:linear+Gaussian}) which encodes the expected asymptotics
clearly becomes dominant.

%%%%%%%%%%%%%%%%%%

\subsubsection{Long-tail realizations}

We now constrain the position of the lead particle to
be larger than $\Xmin$, chosen far from its expectation value, but
in the scaling region.
One may think of different prescriptions for the choice of $\Xmin$.
We have decided to set the value of the diffusion factor,
the width of which
defines the size of the scaling region
(see Eq.~(\ref{eq:shape_traveling})), to a definite value $\alpha$.
Namely:
\be
\exp\left({-\frac{(X_{\text{min}}-m_T)^2}{2\gamma_0 v''(\gamma_0)T}}\right)
\equiv\alpha.
\label{eq:alpha}
\ee
The constant~$\alpha$ has to be picked such that $X_{\text{min}}-m_T$
may be large,
and at the same time, $X_{\text{min}}$ sit well inside
of the scaling region. This means that $\alpha$ must be close to~1,
but not too close to allow for a wide-enough region ahead
of the position $m_T$ for finite~$T$.
We have tested different values, and
eventually picked $\alpha=0.9$ to get the main results discussed
in this paper. The distance $\Xmin-m_T$, which gives the order of magnitude
of the maximum value of $\Delta x$ for which our numerical calculation
may be expected to turn out close to the large-$T$ asymptotics,
are shown in Tab.~\ref{tab:size_scaling_region} for different values of~$T$.
\begin{table}[ht]
  \begin{center}
    \begin{tabular}{cccccc}
      \hline
      $T$ & 1000 & 2000 & 4000 & 8000 & 16000 \\
      \hline
      $\Xmin$ & 1400 & 2800 & 5600 & 11185 & 22350\\
      $X_{\text{min}}-m_T$ & 14.23 & 20.12 & 28.46 & 40.25 & 56.92\\
      \hline
  \end{tabular}
  \end{center}
  \caption{\label{tab:size_scaling_region}
    {\small
      Evolution times $T$ and values of the miminum position $\Xmin$ of the marked
      particle at $T$ used for our numerical calculations.
      $\Xmin$ is set so that $\alpha\simeq 0.9$. 
    In each case, $\Xmin-m_T$ is also shown: It
    represents the typical
    maximum value of the size $\Delta x$ for which
    the interval $[\Xmin-\Delta X,\Xmin]$
    may be considered included in the scaling region.
    }
    }
\end{table}

In each realization, we count the number of particles
between the position $X$ of the rightmost one, and the position $X-\Delta x$.
$X$ and $\Xmin$ do not necessarily coincide, and in general, they do not.
But the ensemble of events is dominated by realizations in
which the rightmost particle is at a distance of order unity to
the right of the position $\Xmin$,
because the events in which $X$ is far from $\Xmin$ are probabilistically
disfavored: Indeed, they are suppressed by a factor of the order
$e^{-\gamma_0(X-\Xmin)}$ if $X$ is still in the scaling region, and
by an additional Gaussian factor if $X$ gets outside of the scaling
region. Since we always count the particles in an interval of fixed
size from the rightmost particle, the particle numbers have no
reason to get
enhanced in the realizations is which $X$ is larger than usual.
Therefore, the fact that the position of the lead particle
is not exactly fixed cannot affect
significantly the observables studied in this paper.

%%%%%%%%%%%%%%%

\paragraph{Probability density.}

\begin{figure}[ht]
  \begin{center}
    \includegraphics[width=\textwidth]{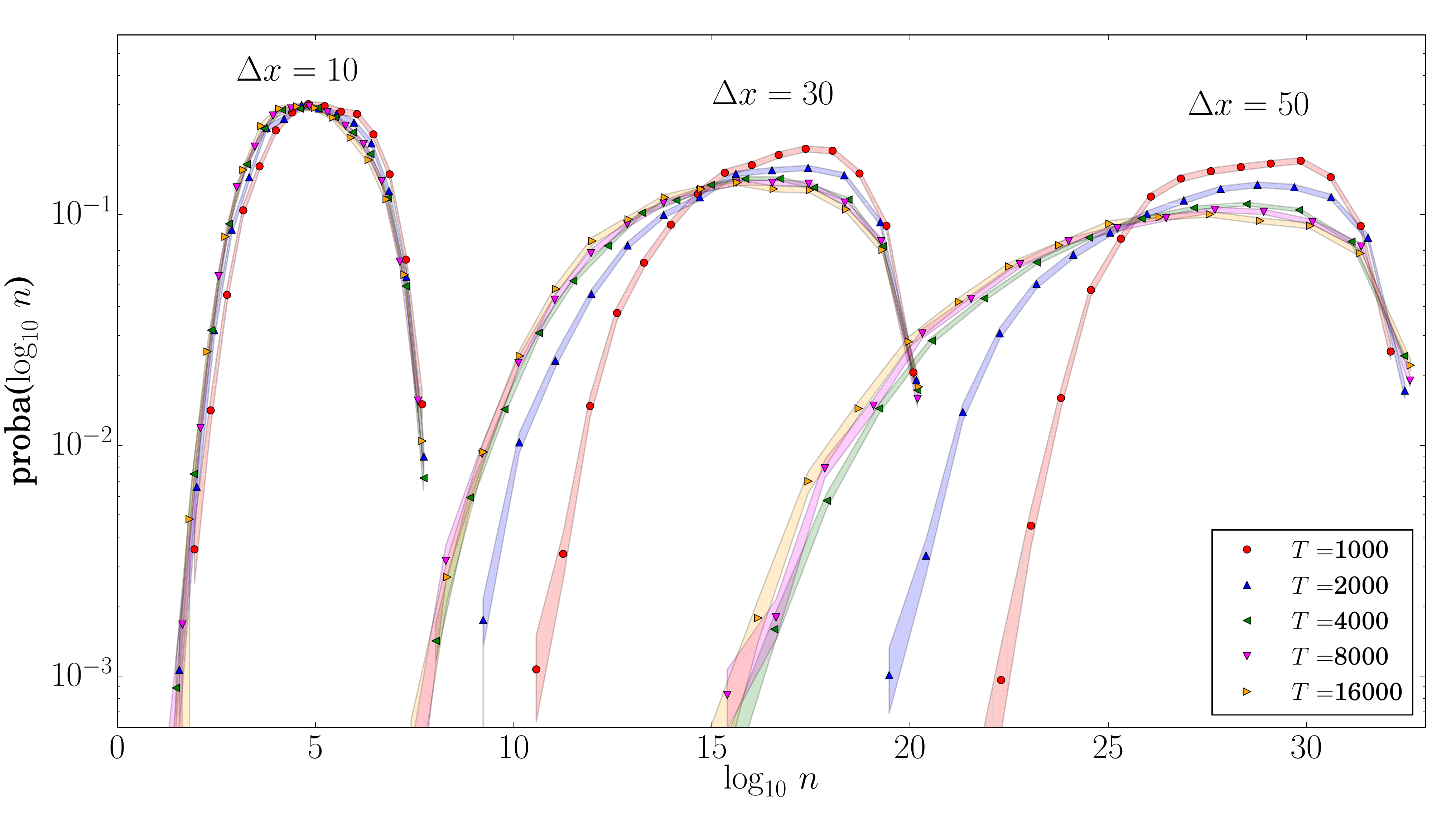}
  \end{center}
  \caption{\label{fig:probability_different_T}
    {\small
      Probability distribution of the number $n$ of particles
      in an interval of size 
      $\Delta x$ for different values of $\Delta x$ and of
      the evolution
    time $T$. The lower bound $\Xmin$ on the position of the lead
    particle is chosen in such a way that $\alpha\simeq 0.9$,
    see Tab.~\ref{tab:size_scaling_region}.}
  }
\end{figure}

The distribution of the particle number $p_n(\Delta x)$
is shown in Fig.~\ref{fig:probability_different_T},
for three different values of $\Delta x$,
for different values of the final evolution
time $T$. For each choice of $T$, $\Xmin$ is set as shown
in Tab.~\ref{tab:size_scaling_region}.

First, we observe that when $T$ gets large,
the distributions converge to an asymptotic
shape for all values of $\Delta x$.
The convergence is faster for small interval sizes
$\Delta x$.

\begin{figure}[ht]
  \begin{center}
    \includegraphics[width=\textwidth]{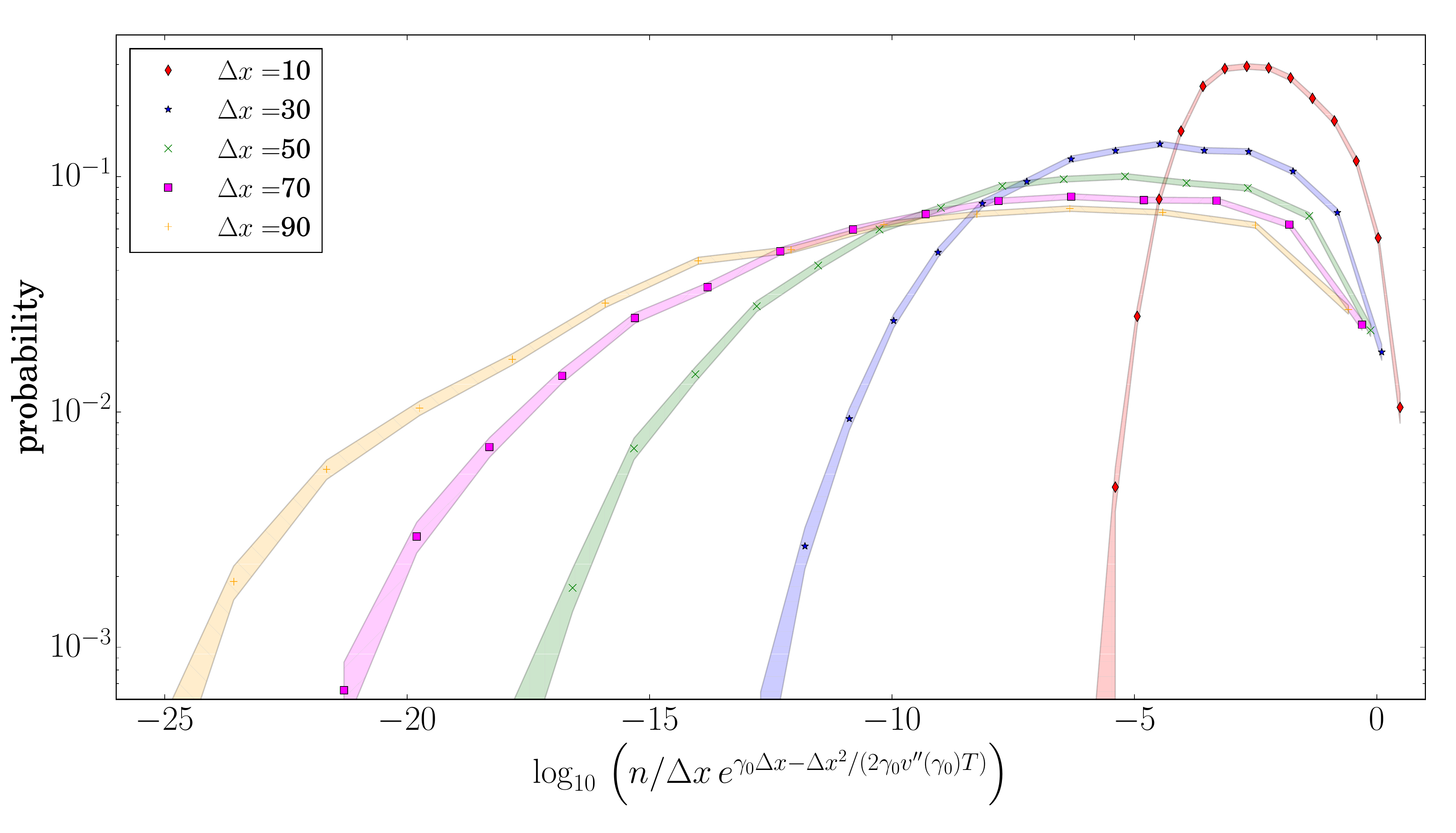}
  \end{center}
  \caption{\label{fig:fixed_T}
    {\small
      Probability distribution of the number $n$ of particles
      in an interval of size 
      $\Delta x$ for $T=16000$ and different values of $\Delta x$; Compare
      to the equivalent plot for the
      unconditioned case displayed in Fig.~\ref{fig:unbiased_proba}b.
      The $x$-axis is shifted by the same $\Delta x$-dependent term (corrected
      by a $T$-dependent term, quite insignificant here)
      as in the latter figure, although in the present case,
      it does not match exactly with the logarithm of
      the mean value of $n$ we expect; See Eq.~(\ref{eq:mean_conditioned}).
  }}
\end{figure}

Second, for asymptotic values of $T$, which seem to be well-approached for
the displayed values of $\Delta x$ as soon as $T\sim 10^4$,
the width of the distribution $p_n(\Delta x)$
clearly increases with $\Delta x$,
at sharp variance with $p_n^{(0)}(\Delta x)$ (see Fig.~\ref{fig:unbiased_proba}).
This increase is particularly striking for the largest value of $T$ for which
we collected numerical data, namely $T=16000$; See Fig.~\ref{fig:fixed_T},
where the probability distribution of the particle number for
different values of $\Delta x$ is displayed, up to $\Delta x=90$.
(Note however that for the largest values of $\Delta x$, the large-$T$
asymptotics have probably not been exactly reached.)

We shall now compute numerically the mean and typical values of this distribution.

%%%%%%%%%%%%%%%

\paragraph{Mean and typical values.}

\begin{figure}[ht]
  \begin{center}
    \includegraphics[width=\textwidth]{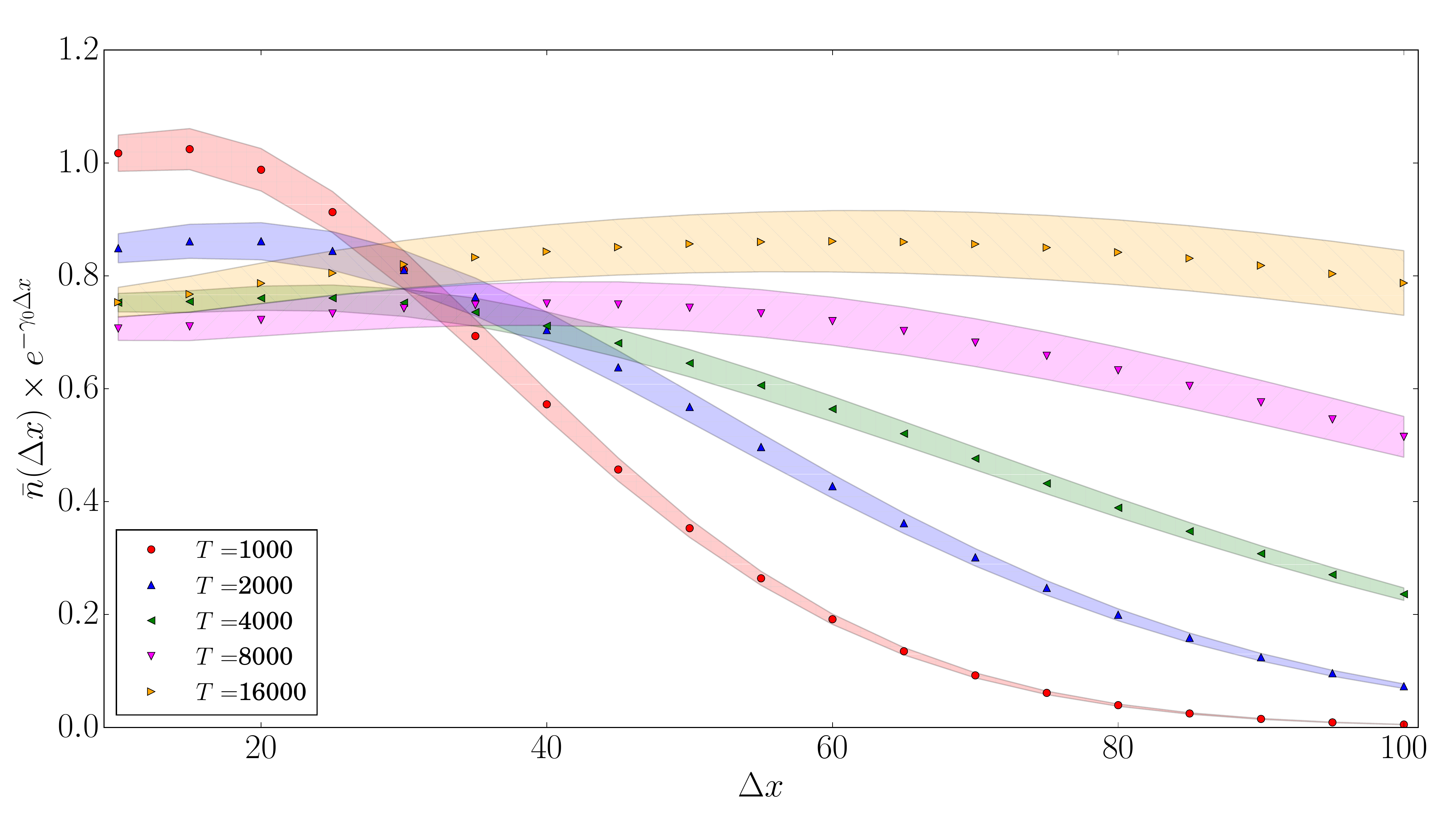}
  \end{center}
  \caption{\label{fig:mean_different_T}
    {\small
    Mean value of the particle number as a function
    of the interval size $\Delta x$, for different values of the evolution
    time $T$. The lower bound $\Xmin$ on the position of the lead
    particle is chosen in such a way that $\alpha\simeq 0.9$, see
    Eq.~(\ref{eq:alpha}) and Tab.~\ref{tab:size_scaling_region} for
    the actual numerical values.}    
  }
\end{figure}

The numerical results for the expectation value of the
particle number $\bar n(\Delta x)$
rescaled by $e^{-\gamma_0\Delta x}$
are shown in Fig.~\ref{fig:mean_different_T} as a function of $\Delta x$
for different values of the final evolution time~$T$.
The statistical uncertainties are again evaluated using the jackknife method.

We see that this rescaled expectation value is consistent with a constant independent
of $\Delta x$ for large times (see the data for $T=16000$).
This data is very different from the one in the unconstrained case;
compare to Fig.~\ref{fig:unbiased_mean}.

\begin{figure}[ht]
  \begin{center}
    \includegraphics[width=\textwidth]{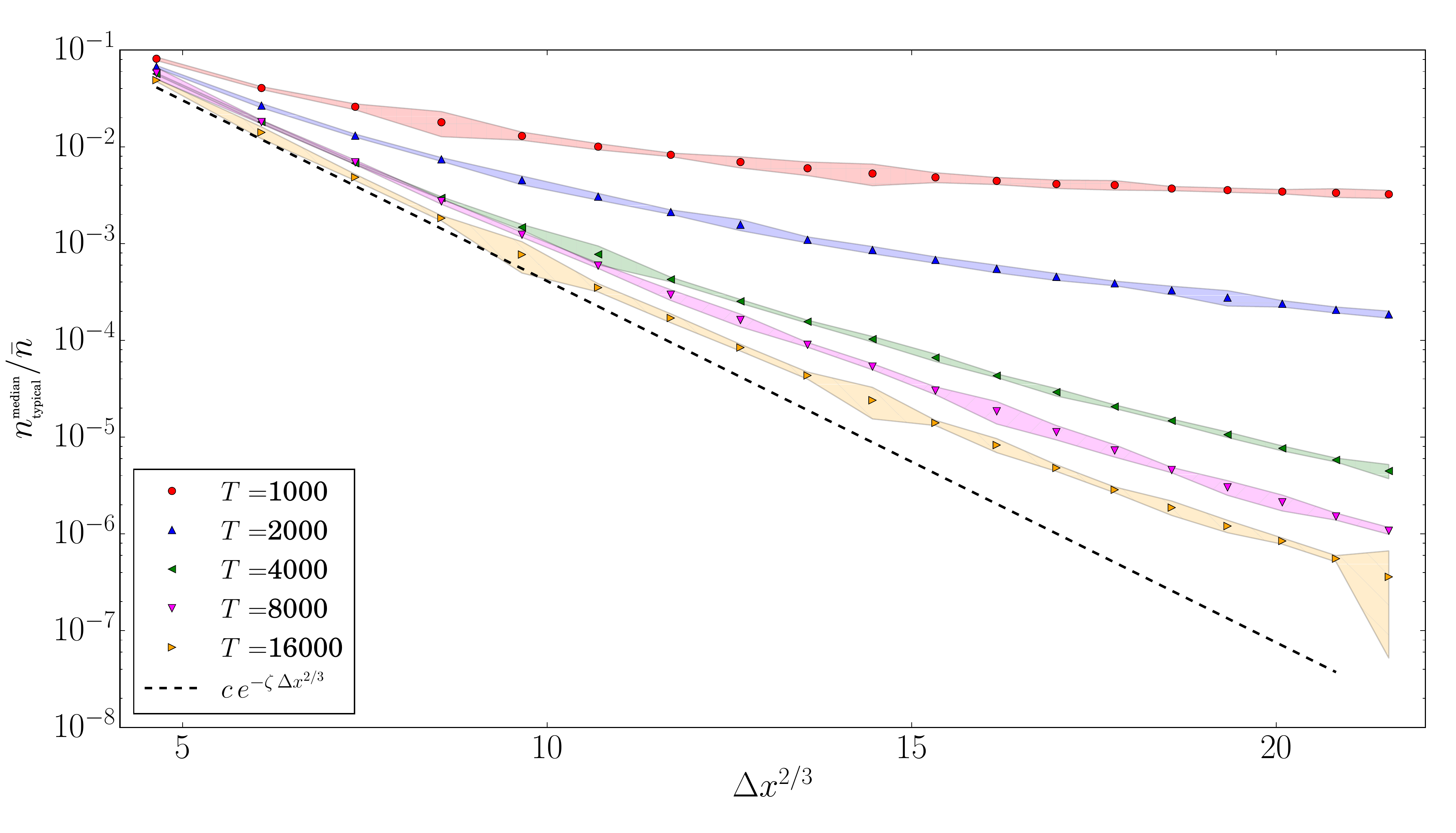}
  \end{center}
  \caption{\label{fig:different_Ta}
    {\small
      Ratio of the typical value of the particle number in the interval
    of size $\Delta x$ to the mean value as a function
    of $\Delta x^{2/3}$, for different values of the evolution
    time $T$.
    The theoretical expectation is shown in dashed line, with the free
    parameters $c$ and $\zeta$, for which no analytical formula is known
    to date, set arbitrarily (see the main text).
    In this plot,
    the typical value is taken to be the median,
    $n^{\text{median}}_{\text{typical}}$ (Eq.~(\ref{eq:median})).
    The dashed straight line represents the function $c\,e^{-\zeta\Delta x^{2/3}}$.
    }
    }
\end{figure}

\begin{figure}[ht]
  \begin{center}
    \includegraphics[width=\textwidth]{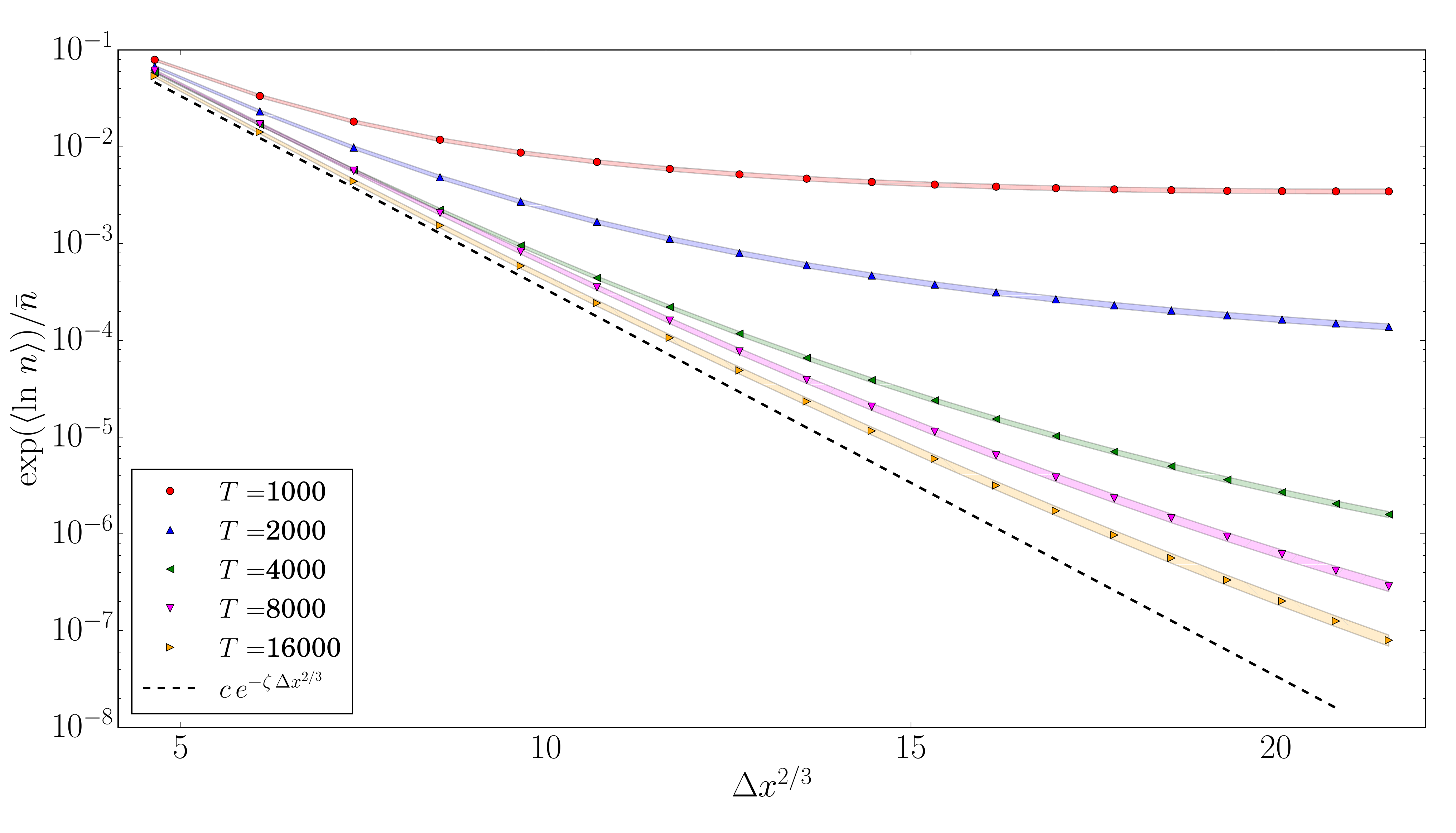}
  \end{center}
  \caption{\label{fig:different_Tb}
    {\small
      The same as in Fig.~\ref{fig:different_Ta}, but
      the typical value is now
      taken to be the exponential of the mean logarithm of the
      particle number, $n^{\text{log}}_{\text{typical}}\equiv e^{\langle\ln n\rangle}$
      (Eq.~(\ref{eq:explog})).
    }
    }
\end{figure}

The numerical results for the typical values of the particle numbers
are shown in Figs.~\ref{fig:different_Ta} and~\ref{fig:different_Tb}.
We use the two estimators described
in Sec.~\ref{sec:typical}.
We plot the ratio of the latter to the expectation values
of the particle number,
on a logarithmic scale, as a function of $\Delta x^{2/3}$.

We see a convergence to a straight line as the evolution time $T$ is taken larger.
This is perfectly consistent with the expected behavior~(\ref{eq:n_typical}), namely
$c\,e^{-\zeta\Delta x^{2/3}}$.
The parameters $c$ and $\zeta$ are set\footnote{
  We have not attempted a regression: Some knowledge of the form of the
  subleading finite-$T$ corrections would be required, which we do not
  have at this point.
}
to $c=2.2$ and $\zeta=0.86$
when the typical value is defined to be the median (Eq.~(\ref{eq:median})),
and
$c=3.3$ and $\zeta=0.92$ when it is defined to the exponential
of the expectation value
of $\ln n$ (Eq.~(\ref{eq:explog})).
Subleading corrections such as a linear prefactor $\propto\Delta x$
are not favored by the data. However, it is difficult to make more
precise statements beyond the leading behavior,
based on these finite-$T$ numerical results.

We note that finite-time (namely finite size of the scaling region) effects are important,
and it is technically not possible to go much beyond $T={\cal O}(10^4)$.
However, it is clear that our numerical results
are consistent with the analytical form that was predicted
in Ref.~\cite{Mueller:2019ror}.

%%%%%%%%%%%%%%%%%%%%%%%%%%%%%%%%%%%%%%%%%%%%%%%%%%%%%%%%%%%%%%%

\section{Conclusions and outlook}

We have designed an algorithm to efficiently
generate realizations of branching random walks
in which there is at least one particle to the right of some predefined
position $\Xmin$
at the final time $T$ at which one analyzes the set of particles.
Using an implementation of this algorithm, we have produced ensembles
of such realizations for which $\Xmin$ was
chosen in the scaling region, far to the right of the mean position $m_T$
of the rightmost particle.
We have focused on the probabilities 
of the particle numbers~$n$ in an interval of variable size $\Delta x$
to the left of the rightmost lead particle.
The main result of this work is that
we have been able to confirm a recent calculation of the mean number of
particles $\bar n$, and a conjecture on the typical
number of particles $n_\text{typical}$. They
are seen to grow like
\be
\boxed{
  \bar n(\Delta x)\propto e^{\gamma_0\Delta x}
  }\quad
  \text{(see Fig.~\ref{fig:mean_different_T})}
  \quad\text{and}\quad
  \boxed{n_\text{typical}\propto e^{\gamma_0\Delta x-\zeta\Delta x^{2/3}}}
  \quad \text{(see Fig.~\ref{fig:different_Ta} or~\ref{fig:different_Tb})}
\ee
with the size $\Delta x$ of the interval, when both the evolution
time $T$ and $\Delta x$ are large.

This numerical result gives more motivation to look for a rigorous proof
of these conjectures and for a formula for the probability distribution
itself (Fig.~\ref{fig:probability_different_T} and~\ref{fig:fixed_T}),
or at least for a complete analytical calculation in some simplified model.
In parallel, a physical picture of the formation of these long tails would
be interesting to develop, 
with the help of more specific Monte Carlo measurements.

The same algorithm could also be used to evolve unbiased realizations to large
times: It is enough to pick $\Xmin$ to the left of $m_T$ by a  distance
chosen in such a way
that the probability to have no particle to the right of $\Xmin$ be
negligible.
We could compare the distribution and its moments in the case
when $\Xmin$ is far out in the scaling region, with
same quantities in the case of an unconstrained boundary. In the latter
case, the shape of the asymptotic distribution is independent of $\Delta x$
(see Fig.~\ref{fig:unbiased_proba}), and is consistent with
a Gumbel distribution.
The mean and typical values of the particle number are proportional to each
other, and consistent with the asymptotic form
$\bar n^{(0)}(\Delta x)\propto\Delta x\,e^{\gamma_0\Delta x}$;
see Fig.~\ref{fig:unbiased_mean}.

A few other observables would be very interesting to measure numerically, with
the help of our algorithm.
For example, the statistics of the branching time of the last common ancestor of
sets of particles chosen in the vicinity of the tip
of the BRW according to given rules: For example,
all particles to the right of a predefined position, or
the $k$ rightmost, or a pair of particles chosen
according to an appropriate weight function of their position.
The considered realizations may be either
unbiased, or conditioned to
have their tip far from its expected position. Analytical
predictions are available in some of these cases \cite{0295-5075-115-4-40005}.

Another outstanding observable would be
the mean distance between neighbouring particles possessing
a predefined order number from
the lead particle, as a function of the latter:
This quantity has been predicted in the case of the branching
Brownian motion with an unconstrained lead particle~\cite{Brunet_2009}.
It would be interesting to measure this observable in an
(effectively) unconditioned Monte Carlo
and, again, investigate how it changes when the position of the rightmost
particle is conditioned to be far from its expected value.
But an implementation
of the continuous BBM is very challenging. We leave this for
future investigations.

\section*{Acknowledgements}
We thank Alexandre Lazarescu for collaborating in an early stage
of this work, and in particular for his help in designing the algorithm.
We also acknowledge his reading of the manuscript.
We thank Satya Majumdar, Philippe Mounaix, St\'ephane Peign\'e,
Gr\'egory Schehr for useful and stimulating discussions.
We are grateful to the CPHT computer support team, whose work
made possible the numerical calculations presented here, performed
on the ``Hopper'' cluster of the PHYMAT mesocenter.
The work of DLA and SM is supported in part by the Agence Nationale
de la Recherche under the project ANR-16-CE31-0019.
The  work  of  AHM  is  supported in part by
the U.S. Department of Energy Grant
DE-FG02-92ER40699.

\section*{Note added (June 29th, 2020)}

A flaw in our algorithm has been brought to our attention,
which may have introduced a bias in the probability distribution of
the realizations we have generated.

After having corrected the algorithm~\cite{BLMM} and reimplemented it,
we have checked that the numbers obtained for the observables
calculated with the new code are indistinguishable from
the ones shown in this paper, in the case $T=1000$ and $\Xmin=1400$.
In view of these first results, we anticipate that the conclusions we have
drawn on $p_n(\Delta x)$ should eventually remain
unchanged.

We will release a new preprint~\cite{NEW}
when all the data, also for higher values of $T$,
has been generated using the corrected code.

We warmly thank \'Eric Brunet for pointing out the problem to us,
and for his prompt help in looking for a fix.

\end{document}